\documentclass[useAMS,usenatbib,usegraphicx]{mn2e}
\usepackage{graphicx,times,epsfig,rotating}

\def\mnras{MNRAS}
\def\aj{AJ}
\def\aap{A\&A}
\def\apj{ApJ}
\def\apjl{ApJ}
\def\apjs{ApJS}
\def\araa{ARA\&A}
\def\pasp{PASP}
\def\nat{Nature}

\def\aaps{A\&AS}

\newcommand{\kmsend}{\mbox{km s$^{-1}$}}
 \newcommand{\kms}{\mbox{km s$^{-1}$ }}

\newcommand{\hpc}{\mbox{h$^{-1}$pc }}

\newcommand{\msun}{\mbox{M$_{\sun}$ }}
\newcommand{\msunend}{\mbox{M$_{\sun}$}}
\newcommand{\lsun}{\mbox{L$_{\sun}$ }}

\newcommand{\lir}{\mbox{L$_{\rm IR}$}}

\newcommand{\cmthree}{\mbox{cm$^{-3}$}}
\newcommand{\cmtwo}{\mbox{cm$^{-2}$}}
\newcommand{\msunyr}{\mbox{M$_{\sun}$yr$^{-1}$ }}
\newcommand{\msunyrend}{\mbox{M$_{\sun}$yr$^{-1}$}}
\newcommand{\htwo}{\mbox{H$_2$}}

\newcommand{\z}{\mbox{$z$}}

\newcommand{\zsim}{\mbox{$z\sim$ }}

\newcommand{\sunrise}{\mbox{\sc sunrise}}
\newcommand{\gadget}{\mbox{\sc gadget-3}}
\newcommand{\gadgettwo}{\mbox{\sc gadget-2}}

\newcommand{\starburst}{\mbox{\sc starburst99}}
\newcommand{\mappings}{\mbox{\sc mappingsiii}}
\newcommand{\turtlebeach}{\mbox{\sc turtlebeach}}

\newcommand{\xco}{\mbox{$X_{\rm CO}$}}
\newcommand{\xcounits}{\mbox{cm$^{-2}$/K-km s$^{-1}$}}
\newcommand{\qeos}{\mbox{$q_{\rm EOS}$}}
\title[\xco \ in Discs and Mergers]{The CO-\htwo \ Conversion Factor in Disc
  Galaxies and Mergers}

\author[Narayanan et al]{Desika\, Narayanan$^{1}$\thanks{E-mail:
    dnarayanan@as.arizona.edu}\thanks{Bart J. Bok Fellow}, Mark
  Krumholz$^{2}$, Eve C. Ostriker$^3$, Lars Hernquist$^4$\\$^{1}$Steward
  Observatory, University of Arizona, 933 N Cherry Ave, Tucson, Az,
  85721\\$^{2}$Department of Astronomy and Astrophysics, University
  of California, Santa Cruz, CA, 95064\\$^{3}$Department of
  Astronomy, University of Maryland, College Park,
  20742\\$^{4}$Harvard-Smithsonian Center for Astrophysics, 60 Garden
  St., Cambridge, Ma 02138\\}
\begin{document}

\date{Accepted by MNRAS}

\pagerange{\pageref{firstpage}--\pageref{lastpage}} \pubyear{2010}

\maketitle

\label{firstpage}

\begin{abstract}

Relating the observed CO emission from giant molecular clouds (GMCs)
to the underlying \htwo \ column density is a long-standing problem in
astrophysics.  While the Galactic CO-\htwo \ conversion factor (\xco)
appears to be reasonably constant, observations indicate that \xco
\ may be depressed in high-surface density starburst
environments. Using a multi-scale approach, we investigate the
dependence of \xco \  on the galactic environment in numerical
simulations of disc galaxies and galaxy mergers.  \xco \ is
proportional to the GMC surface density divided by the integrated CO
intensity, $W_{\rm CO}$, and $W_{\rm CO}$ is related to the kinetic
temperature and velocity dispersion in the cloud.  In disc galaxies
(except within the central $\sim$ kpc), the galactic environment is
largely unimportant in setting the physical properties of GMCs
provided they are gravitationally bound.  The temperatures are roughly
constant at $\sim10$ K due to the balance of CO cooling and cosmic ray
heating, giving a nearly constant CO-\htwo \ conversion factor in
discs.  In mergers, the velocity dispersion of the gas rises
dramatically during coalescence.  The gas temperature also rises as it
couples well to the warm ($\sim50$ K) dust at high densities ($n >
10^4$ \cmthree).  The rise in velocity dispersion and temperature
combine to offset the rise in surface density in mergers, causing \xco
\ to drop by a factor of $\sim2-10$ compared to the disc
simulation.  This model predicts that high-resolution ALMA
observations of nearby ULIRGs should show velocity dispersions of
10$^1$-10$^2$ \kmsend, and brightness temperatures comparable to the
dust temperatures.

\end{abstract}

\section{Introduction}
\label{section:introduction}

Stars form in giant molecular clouds (GMCs) whose primary constituent
is molecular hydrogen, \htwo.  Because \htwo \ lacks a permanent
dipole moment, and the lowest lying excited state capable of
quadrupole emission requires temperatures $\sim 500$ \ K to be excited,
the physical conditions in the cold ($\sim 10$ K) molecular gas are
typically probed via tracer molecules, rather than by direct detection
of \htwo.

Carbon Monoxide ($^{12}$C$^{16}$O; hereafter, CO) is the second most
abundant molecule in GMCs. Because the J=1-0 rotational transition of
CO lies only $\sim5$ K above ground, has a relatively low effective
density ($\sim 10^{2-3}$\cmthree) for excitation \citep{eva99},
and has a wavelength of $\sim3$ mm which is readily observable from
the ground, CO (J=1-0)  has historically been one of the most commonly
used tracers of physical conditions in the molecular ISM.  

A large uncertainty in using CO to trace \htwo \ gas is relating the
observed CO line luminosity to the underlying \htwo \ column density.
However, despite the fact that CO/\htwo \ abundances vary strongly
within GMCs \citep[e.g. ][]{ste95,lee96,hol99,glo10,glo11}, a
multitude of observations suggests that the conversion factor
between CO and \htwo \ is reasonably constant in Galactic GMCs,
following the relation:

\begin{equation}
\label{eq:xco_definition}
\xco = 2-4 \times 10^{20} \rm {cm}^{-2}/(K-\kms)
\end{equation}
where \xco \ is the CO-\htwo \ conversion factor in units of \htwo
\ column density divided by velocity-integrated CO line
intensity\footnote{\xco \ is sometimes referred to in the literature
  as the ``$X$-factor''.  We will use \xco \ and $X$-factor
  interchangeably.}. Lines of evidence for a relatively constant \xco
\ include comparisons between CO luminosities and molecular column
densities determined via a variety of techniques, including dust
extinction \citep{dic75}, $\gamma$-ray emission
\citep{blo86,str96,abd10} and thermal dust emission
\citep{dam01,dra07b}.

%idence for a relatively constant \xco \ in the
%Galaxy comes from a variety of independent techniques including
%extinction measurements to determine the \htwo \ mass \citep{dic75},
%observations of $\gamma$-rays which result from the interaction of
%cosmic rays with the ISM \ as a measure of the underlying neutral gas
%mass \citep{blo86}, and virial mass measurements of GMCs from the
%optically thick CO emission line \citep{sol87}.

Beyond this, the CO-\htwo \ conversion factor appears to have the same
relatively narrow range of values in galaxies in the Local Group as
well \citep{bli07,wol10}, though there may be some variations
associated with metallicity
\citep{wil95,ari96,bos02,ros03,isr05,bel06,ler06,bel07,bol08,ler11}. The
relatively narrow distribution of values for \xco \ in Local Group
GMCs\footnote{\xco \ also appears to be reasonably constant in diffuse
  \htwo \ gas in the Galaxy \citep{lis10}.  This has been attributed
  to the offsetting effects of of lower CO abundances with respect to
  \htwo \ (most of the carbon is in C$^+$) and a large W$_{\rm
    CO}$/N$_{\rm CO}$ ratio in low extinction gas. See the discussion
  in \citet{pet11} for more details. } may arise from the fact that
molecular clouds appear have remarkably similar physical properties in
both the Milky Way and nearby galaxies.  GMCs in the Galaxy and local
Universe appear to have a nearly constant surface density of 85-100
\msun/pc$^{2}$, obey the size-linewidth relation, and have relatively
low kinetic temperatures of 10-20 K
\citep{sol87,bli07,hey09}. Magnetohydrodynamic and radiative transfer
modeling by \citet{glo10}, \citet{glo11} and \citet{she11a,she11b}
have shown that simulated GMCs with mean densities, sizes, velocity
dispersions, and metallicities comparable to those found in the Galaxy
naturally produce \xco \ conversion factors comparable to
Equation~\ref{eq:xco_definition}.

The situation becomes more complex in starburst galaxies.  By
utilising high-spatial resolution interferometric mapping of nearby
(ultra)luminous infrared galaxies ([U]LIRGs;
\lir$>$[$10^{12}$]$10^{11}$ \lsun), \citet{sol97}, \citet{dow98} and
\citet{dow03} have shown that the application of the ``standard''
Galactic \xco \ conversion factor would cause the inferred molecular
gas mass to exceed the dynamical mass in these galaxies.  In this
case, the constraints on \xco \ are in the range $\sim 2-10 \times
10^{19} \rm {cm}^{-2}/\rm{K-km/s}$: a factor of 2-20 lower than the
Galactic value.  Other observational evidence from local starbursts
\citep{hin06,mei10}, the Galactic centre \citep{oka98} and high-\z
\ submillimetre-galaxies \citep{tac08} have all corroborated this
picture that \xco \ may be lower in regions of high molecular surface
density.

The exact origin of a lower \xco \ factor in starburst galaxies is not
entirely clear. Models by \citet{mal88} which predate the
aforementioned observations, predicted that warmer molecular gas
temperatures in infrared-luminous galaxies may drive a lower \xco
\ conversion factor due to an increase in CO brightness temperature
with kinetic temperature in optically thick clouds.  Alternatively,
\citet{dow98} suggest that the CO line width in starbursts traces a
combination of the gaseous and stellar potential, rather than just the
\htwo \ mass.  In the case where CO is optically thick, the observed
velocity-integrated CO line intensity can increase with the velocity
dispersion. \citet{mal88} and \citet{she11b} also postulated a similar
effect if the CO linewidths were larger than their typical virial
values.

While the scaling of \xco \ with environmental parameters is not yet
known, the ramifications are profound.  For example, if \xco \ does
indeed systematically vary in higher surface density environments, our
current understanding of the normalisation and index of the
Kennicutt-Schmidt star formation rate-gas surface density relation in
star forming galaxies may change \citep{ken98a,dad10b,gen10}.  At
higher redshifts, as high gas surface density galaxies begin to
contribute substantially to the cosmic star formation rate density
\citep[e.g. ][]{lef05, hop10, hop10b}, the variation of \xco \ with
environment may affect observed values of the cosmic evolution of
$\Omega_{\htwo}$.  More generally, the interpretation of forthcoming
results from the EVLA and ALMA will be severely crippled without an
understanding for how to relate the observed CO line flux to the
quantity of interest: \htwo \ gas mass.

In this area, numerical simulations can offer some guidance. Indeed,
understanding the origin of \xco \ in galaxies is difficult in
that many of the physical parameters driving the relation are
coupled.  The CO-\htwo \ conversion factor, \xco, has dimensions of:
\begin{equation}
\xco \propto \Sigma/(W_{\rm CO}) \propto \Sigma/(\sigma \times T_{\rm B})
\label{equation:xco_units}
\end{equation}
where $W_{\rm CO}$ is the velocity-integrated CO intensity, $\Sigma$
is the gas surface density, $T_B$ is the brightness temperature of the
line, and $\sigma$ is the velocity dispersion.  To first order, $T_B$
is related to the kinetic temperature of the gas ($T_{\rm k}$) when
the line is thermalised.  However, the kinetic temperature of the gas
can depend on molecular abundances, gas densities, dust temperatures
and the background radiation field \citep[e.g. ][]{nar06b,kru11a}.
The same physical processes which can cause changes in these
parameters may also cause the GMC surface densities to change as well.
The problem is well-suited for numerical simulations.

 Building on the seminal work of \citet{mal88} and more recent
 simulations of \citet{glo10,glo11} and \citet{she11a,she11b}, we present
 the first models investigating the CO-\htwo \ conversion factor in
 hydrodynamic simulations of isolated disc galaxies and disc galaxy
 mergers. This is the first paper in a series.  In this work, we aim
 to understand whether \xco \ varies between ``normal'' disc galaxies
 and galaxy mergers, and if so, why.  In order to do this, we couple
 smoothed-particle hydrodynamic simulations of galaxies in evolution
 with dust and molecular line radiative transfer calculations to
 self-consistently calculate the kinetic temperature of and emissivity
 from GMCs in our models.  Our main result is that higher kinetic
 temperatures and velocity dispersions in the GMCs naturally arise
 during mergers, and contribute to lower values of \xco \ in these
 systems.

Our paper is organised as follows: In \S~\ref{section:methods}, we
present our radiative transfer and hydrodynamics methodology; in
\S~\ref{section:observationalproperties}, we discuss the synthetic
observational and physical properties of our model galaxies in an
effort to aid comparisons to observations; in
\S~\ref{section:results}, we discuss how the CO-\htwo \ conversion
factor varies in disc galaxies and mergers in our models; in
\S~\ref{section:discussion}, we discuss the implications of our
findings, and in \S~\ref{section:summary} we summarise our results.

\section{Methods}
\label{section:methods}

Our goal is to simulate the emission from GMCs on galaxy-wide
scales. This involves simulating galaxies in evolution, the physical
state of GMCs, molecular line radiative transfer through the clouds,
and dust and molecular line radiative transfer through the galaxy.  In
this section, we describe these simulations, and the relevant
assumptions that go into our modeling.  This involves combining a
large number of simulation codes.  In light of this, to guide the
reader through the numerical details and equations in this section, we
first summarise them more generally here.

We first simulate the hydrodynamic evolution of both disc galaxies and
mergers.  It is from these simulations that we know the global
distribution of stars, gas and metals in the galaxy, and their
physical properties.  The radiative transfer occurs in
post-processing.  We project the physical conditions of the particles
onto an adaptive mesh using the SPH smoothing kernel. The base mesh is
5$^3$ spanning a 200 kpc box.  The cells refine recursively into 2$^3$
subcells based on the refinement criteria the relative density
variations of metals ($\sigma_{\rho m}/<\rho_m>$) should be less than
0.1, and the $V$-band optical depth across a cell be less than unity.
The maximum refinement level was 11, such that the smallest cells in
this mesh are of order $\sim70$ pc across.

The surface density of and velocity dispersion within the GMCs are set
by the physical conditions in the hydrodynamic galaxy evolution
simulations.  A subgrid prescription comes into play when GMCs are
unresolved (i.e. when cells in the adaptive mesh are very large). We
assume that all of the \htwo \ mass in the cell is in the GMC and we
calculate the HI-\htwo \ balance via analytic models (described
below).  From this, the complete physical conditions (except for the
temperature) of the GMCs are described by the hydrodynamic
galaxy evolution simulations.  The temperatures of the clouds are
calculated by assuming thermal equilibrium between gas heating (by the
grain photoelectric effect and cosmic rays), gas cooling (via
molecular and atomic line cooling), dust heating (from the ambient
radiation field), thermal dust cooling, and some energy exchange between
gas and dust.

With the physical properties of the galaxies and GMCs known, we then
proceed to calculate the emergent CO emission from the clouds.  We
calculate the CO line emission from the GMCs utilising an escape
probability formalism.  The radiation from these clouds then interacts
with other clouds in the galaxy, and the level populations of CO are
calculated by the balance of radiative absorptions, stimulated
emission, spontaneous emission, and collisions with \htwo \ and He.

At this point, the general reader should be equipped to understand the
general results of this paper.  For the remainder of this section, we
elaborate on this abbreviated description.  Throughout, we assume
$h=0.7$.

\subsection{Smoothed Particle Hydrodynamic Simulations of Galaxies in Evolution}
\label{section:sph}
We simulate the hydrodynamic evolution of both idealised isolated disc
galaxies, and mergers between these discs.  The purpose of the
hydrodynamic simulations is to calculate the spatial distribution of
the neutral ISM, stars and metals.  It is from the neutral ISM that we
will calculate the molecular gas properties, and, as we will discuss,
the radiation from the stars and dust in the metals that determine the
IR radiation field.  Here, we describe the components of the model
most pertinent to this study, namely the physics of the ISM and star
formation prescriptions.  For a more full understanding of the
underlying algorithms in \gadget, please refer to
\citet{spr02,spr03a,spr05b} and \citet{spr05a}\footnote{We note that
  \citet{spr05a} describes the publicly-available \gadgettwo, whereas
  the work in this paper utilises \gadget, a non-public modified
  version of \gadgettwo.  The main improvement in \gadget \ over
  \ \gadgettwo \ is better load balancing on parallel processors.}.

The galaxies are simulated with a modified version of the publicly
available SPH code, \gadget \ \citep{spr05b}.  The ISM is modeled as
two-phase, with cold clouds embedded in a hot, pressure-confining
medium \citep{mck77, spr03a}.  Numerically, this is realised via
hybrid SPH particles.  The cold gas mass grows via radiative cooling
of the hot phase, and cold gas is converted to hot gas through the
heating associated with star formation.

Stars form in the cold ISM according to a relation SFR $\propto
\rho_{\rm cold}^{1.5}$. The normalisation of this relation is set in order
to match the local $\Sigma_{\rm SFR}-\Sigma_{\rm gas}$ relation
\citep{ken98a,ken98b,spr00,cox06a}.  

Supernova pressurisation of the ISM is modeled via an ``effective''
equation of state \citep[see Figure 4 of ][]{spr05a}.  Here, we assume
a modest pressurisation of $q_{\rm EOS}=0.25$ in the \citet{spr05a}
formalism.  This corresponds to a mass-weighted ISM temperature of
$\sim10^{4.5}$K.  In the Appendix we relax the star formation and
equation of state assumptions in order to test the validity of our
results.

The simulations here are not cosmological: the discs are set up in an
idealised manner in order to maximise spatial resolution.  Here, the
gravitational softening length for baryons is 100 \hpc, and 200 \hpc
for dark matter.  The discs are initialised according to the
\citet{mo98} formalism, and are bulgeless.  They are embedded in dark
matter halos with \citet{her90} density distributions.  

In order to compare with observations in a meaningful manner, we aim
to simulate galaxies comparable to those found in the local Universe.
Accordingly, our isolated discs are initialised inside haloes of mass
$\sim1.9\times 10^{12}$ \msunend, baryonic mass of
$\sim8\times10^{10}$ \msunend, circular velocity of 160 \kmsend, and
with 40\% of the baryons in the form of gas.  

The mergers are binary 1:1 mergers between discs constructed in the
same manner.  We simulate three mergers of slightly higher mass in
order to ensure that they undergo a luminous starburst comparable to
the most extreme ones seen in the local Universe ($\sim100
\ \msunyrend$).  In particular, the discs that comprise the binary
mergers have a rotation speed of 225 \kmsend, halo mass of $\sim 5
\times 10^{12}$ \msunend, and baryonic mass of $\sim 2.2 \times
10^{11}$ \msunend.  The mergers are set on an orbit with angles
($\theta_1,\phi_1,\theta_2,\phi_2$) = (30,60,-30,45),
(-109,-30,71,-30) and (0,0,0,0). The angles for the first two orbits
are arbitrary, and were chosen to represent relatively ``normal''
orbits in our library of simulations.  The last merger is a coplanar
one, and represents an extreme starburst with an extended duration,
which we include simply for comparison.  We choose the first merger as
our ``fiducial'' merger for the remainder of this paper as this
particular model is well-studied in the literature\footnote{In
  \S~\ref{section:observationalproperties}, we discuss the physical
  and simulated observational properties of our fiducial merger to
  highlight its similarity to observed local galaxies.}, and focus
particularly on the snapshot when the star formation rate is at its
peak.  The results from all simulations are similar, and we discuss
the minor differences that do exist when necessary.

\subsection{Physical Properties of Giant Molecular Clouds}

We assume that the entire neutral mass in a given cell is locked in a
cloud which is spherical, isothermal, and of constant density.  We
determine the surface density of the neutral gas via
\begin{equation}
\Sigma_{\rm cloud} = {\rm max}( \Sigma_{\rm cell},100 \ \msun {\rm pc}^{-2})
\end{equation}
where $\Sigma_{\rm cell}$ \ is the surface density of the cell in the
SPH simulation.  In this model, when the cloud is resolved, we use the
surface density as calculated in the simulations.  When the cloud is
unresolved, we adopt a subresolution surface density comparable to
observed values of GMCs \citep[e.g. ][]{sol87,bli06}.

We then determine the \htwo \ fraction of the neutral ISM utilising
the analytic formalism of \citet{kru08, kru09a} and \citet{mck10}.  This
prescription aims to model the balance between the dissociation of
molecules by Lyman-Werner band photons, and the formation of molecules
on dust grains.  We refer the readers to the aforementioned papers for
the full derivation, and simply repeat the numerical prescription
here.  The molecular fraction is given by:
\begin{equation}
\label{eq:kmt}
f_{\rm H2} \approx 1 - \frac{3}{4}\frac{s}{1+0.25s}
\end{equation}
for $s<2$ and $f_{\rm H2} = 0$ for $s\geq 2$.  $s = {\rm ln}
(1+0.6\chi + 0.01\chi^2)/(0.6\tau_{\rm c})$, where $\chi =
0.76(1+3.1Z'^{0.365})$, and $\tau_{\rm c} = 0.066\Sigma_{\rm
  cloud}/(\msun {\rm pc^{-2}})\times Z'$.  $Z'$ is the metallicity
divided by the solar metallicity.  This formalism for deriving
$f_{\rm H2}$ assumes chemical equilibrium.  

It is worth a quick note that there are numerous prescriptions for
determining the HI/\htwo \ balance in the ISM of simulations, some of
which include time-dependent chemistry.  \cite{bli06} developed an
empirical pressure-based methodology for calculating the \htwo
\ fraction in the neutral ISM, based on observations of local
galaxies.  Similarly, both semi-analytic models \citep{obr09a,obr09b},
as well as full numerical solutions exist which model the effect of
dissociating photons through models of galaxies
\citep[e.g. ][]{pel06,dob08,rob08,pel09,gne09,gne10}.  We motivate our usage
of the analytic prescription of \citet{kru09a} for two reasons.
First, some observational evidence suggests that on small scales
($<100$ pc), Equation~\ref{eq:kmt} may fare better than pressure-based
prescriptions in describing the state of the neutral ISM in
low-metallicity dwarf galaxies \citep{fum10}.  Second, a comparison
between Equation~\ref{eq:kmt} and a numerical treatment of
time-dependent chemical reaction network and radiative transfer in
galaxies suggests that the analytic approximation is reasonable at
metallicities above 0.01 $Z_{\odot}$ \citep{kru11b}.  Because we aim
to model actively star-forming systems in this work, we find that the
mass-weighted metallicity of our model clouds is always higher than
this fiducial value and expect that the analytic approximation is
therefore reasonable.

 With $\Sigma_{\rm cloud}$ and ${M}_{\rm H2}$ defined, the radius of
 the cloud is known. In order to account for the turbulent compression
 of gas, we scale the volumetric densities of the GMCs by a factor
 $e^{\sigma_\rho^2/2}$ where numerical simulations show 
\begin{equation}
\label{eq:turbulentcompression}
\sigma_\rho^2 \approx {\rm ln}(1+3M_{\rm 1D}^2/4)
\end{equation}
where $M_{\rm 1D}$ is the 1 dimensional Mach number\footnote{We note that
  other authors have found a range of possible forms for
  Equation~\ref{eq:turbulentcompression}.  For example, \citet{lem08}
  find $\sigma_\rho^2 \approx 0.6 {\rm ln}(1+0.5M_{\rm 3D}^2)$, while
  \citet{pri11} find $\sigma_\rho^2 \approx {\rm ln}(1+1/9(M_{\rm 3D}^2))$
  where $M_{\rm 3D}$ is the 3D Mach number.}  of the turbulence
\citep[][see also \citet{lem08}]{ost01,pad02}.  Because the
temperature calculation is dependent on the density of the GMC (see
below), solving for the density and temperature simultaneously is a
computationally lengthy process for the multi-million-cell grids that
concern us.  Thus, to calculate the turbulence-driven density
enhancement, we assume the temperature of the GMC is 10 K, which as we
shall show, is a good approximation for the bulk of the GMCs in these
simulations.

 We calculate the 1D velocity dispersion in the cloud:
\begin{equation}
\sigma = {\rm max}( \sigma_{\rm cell},\sigma_{\rm vir})
\end{equation}
where $\sigma_{\rm cell}$ is the mean square sum of the subgrid
turbulent velocity dispersion within the GMC and the resolved
nonthermal velocity dispersion.  The subgrid turbulent velocity
dispersion is calculated from the external pressure from the hot ISM
\citep[][]{rob04} using $\sigma^2=P/\rho_{cell}$ though we impose a
ceiling of 10 \kms which comes from average values found in turbulent
feedback simulations \citep[e.g ][]{dib06,jou09,ost11}. The resolved
nonthermal component is calculated by finding the turbulent velocity
dispersion of the nearest neighbouring cells in the simulation.  In
detail, we calculate the standard deviation of the velocities of the
nearest neighbour cells in the $\hat{x}$, $\hat{y}$ and $\hat{z}$
directions, and define the nonthermal velocity dispersion as the mean
of these.  In cases where the GMC is unresolved, a floor $\sigma_{\rm
  vir}$ is set by assuming the cloud is in virial balance with a
virial parameter $\alpha_{\rm vir} = 1$,
for $\alpha_{\rm vir} \equiv 5 \sigma_{\rm vir}^2 R/(GM)$, so that
\begin{equation}
\label{equation:sigma}
\sigma_{\rm vir} = 2.2 \ \kms \left[\frac{M }{10^5 \  \msun}\right]^{1/4}
\end{equation}
for $\Sigma_{\rm cloud}=100 \ \msun {\rm pc}^{-2}$
where $M$ is the mass of the cloud.  

Finally, we calculate the
temperature of the model GMCs.  The model is based on that developed
by \citet{kru11a}, and we describe the relevant details here as it is
an important aspect of our model. The temperature of the molecular ISM
is determined by a balance of heating and cooling processes in the
gas, heating and cooling of the dust, and a dust-gas thermal exchange.
For the gas, we consider grain photoelectric heating at a rate per H
nucleus $\Gamma_{\rm pe}$, cosmic ray heating at a rate $\Gamma_{\rm
  CR}$, and cooling via either CII or CO line cooling at a rate
$\Lambda_{\rm line}$.  The dust can be heated by the background
infrared radiation field at a rate $\Gamma_{\rm dust}$, and cool via
thermal emission at a rate $\Lambda_{\rm dust}$.  Finally, there is an
energy exchange between dust and gas at a rate $\Psi_{\rm gd}$ where
$\Psi_{\rm gd}$ is positive if the dust is hotter than the gas.  If the gas and
dust are in thermal balance, then we have the following equations:
\begin{eqnarray}
\Gamma_{\rm pe} + \Gamma_{\rm CR} - \Lambda_{\rm line} + \Psi_{\rm gd} = 0\\
\Gamma_{\rm dust} - \Lambda_{\rm dust} - \Psi_{\rm gd} = 0  
\end{eqnarray}
The equation is solved by simultaneously iterating on the temperatures
of the gas and dust\footnote{We note that this dust temperature is
  not always the same as the temperature calculated by \sunrise
  \ (\S~\ref{section:sunrise}).  This makes little difference on the
  final results.  We discuss this in more detail in
the Appendix.}.

The grain photoelectric heating rate is assumed to be attenuated by
half the mean extinction of the cloud (as the heating rate is expected
to decrease toward the cloud interiors) and is given by:
\begin{equation}
\Gamma_{\rm pe} = 4 \times 10^{-26}G_0'e^{-N_{\rm H}\sigma_{\rm d}/2} {\rm erg \ s^{-1}}
\end{equation}
where $G_0'$ is the FUV intensity relative to the Solar neighborhood,
and $\sigma_{\rm d}$ is the dust cross section per H atom to UV
photons.  Here, we assume that the $G_0' = 1$ and $\sigma_d =
1\times10^{-21}$ cm$^{-2}$.  Test models in which we scale $G_0$ by
the star formation rate density compared to that found in the solar
neighbourhood \citep[e.g. ][]{ost10} have similar results to those
presented in this work, and are presented in the Appendix.

The cosmic ray heating rate is given by: 
\begin{equation}
\Gamma_{\rm CR} = \zeta' q_{\rm CR} \ {\rm s^{-1}}
\end{equation}
where $\zeta'$ is the cosmic ray ionisation rate (here, assumed to be
2 $\times 10^{-17} Z' {\rm s}^{-1}$), and $q_{\rm CR}$ is the thermal
energy increase per cosmic ray ionisation.  For \htwo, $q_{\rm CR}
\approx 12.25$ eV \citep[though note that this value is quite
  uncertain; see discussion in Appendix A4 of ][]{kru11a}, and for HI,
$q_{\rm CR} = 6.5$ eV \citep{dal72}.  We utilise a constant cosmic ray
heating rate for all simulations.  Some models suggest that there may
be enhanced cosmic ray fluxes during starbursts which would increase
the \htwo \ gas temperature \citep{pap10a,pap10b}, and further enhance
the effects found in our Results section.

Finally, in a subset of models we have explored the potential effects
of turbulent heating on molecular clouds. In unresolved GMCs we can
estimate this heating rate based on numerical experiments on the rate
of turbulent dissipation: $\Gamma_{\rm turb} \approx 1.5 \times
\sigma^3/R$, where $R$ is the GMC radius \citep{mck07}. For
resolved GMCs, we can measure the turbulent heating rate directly from
the code. Bulk turbulent motions can be converted to heat through two
pathways: adiabatic compression and viscous dissipation. The
compressive heating rate per unit mass is $\Gamma_{\rm comp} = P
(\nabla \cdot \mathbf{v})/\rho$, and we can evaluate this directly
from the density and velocity fields output by Gadget. The viscous
dissipation rate per unit mass is $\Gamma_{\rm visc} =
(\mathbf{\pi}_{\rm visc} \cdot \nabla) \cdot \mathbf{v} / \rho$, where
$\mathbf{\pi}_{\rm visc}$ is the viscous stress tensor. The code
relies on implicit dissipation rather than an explicit viscosity, but
we can estimate the viscous heating rate produced by that implicit
dissipation by noting that the Reynolds number must be $\sim 1$ on the
resolution scale of the code \citep{off09}. This implies that the
dynamic viscosity is $\eta \approx \rho v h$, where $h$ is the SPH
smoothing scale. Given this approximation, the components of the
viscous stress tensor are $\pi_{ij,\rm visc} = \eta [\partial
  v_i/\partial x_j + \partial v_j/\partial x_i - (2/3) \partial
  v_i/\partial x_j \delta_{ij}]$, and which we can again evaluate
directly from the density and velocity fields output by Gadget. We
find that the effects of turbulent heating are modest in both the
resolved and unresolved cases. In our fiducial merger including
viscous dissipation reduces \xco \ by $\sim 30\%$, while the the
fiducial disc it reduces \xco \ by less than a few
percent. Hereafter we neglect this heating term, though we note that
including it would only enhance the results we present below.

%Finally, in a subset of models we have explored the potential effects
%of turbulent and viscous heating on the molecular clouds.  The
%turbulent heating is given by $\Gamma_{\rm turb} = P (\nabla \cdot
%v)/\rho$ for resolved GMCs.  For unresolved GMCs, we assume the rate
%of energy dissipation per unit mass is given by $1.5 \times
%\sigma^3/R$ \citep{mck07}, where $R$ is the radius of the GMC.
%Similarly, for the viscous heating rate, we calculate an energy rate
%per unit volume per unit mass as $\Gamma_{\rm visc} = (\sigma_{\rm
%  visc} \cdot \nabla) \cdot v/\rho$ where $\sigma_{\rm visc}$ is the
%viscous stress tensor.  We estimate the magnitude of the stress tensor
%by assuming a Reynolds number of unity which consequently leads to a
%stress tensor magnitude of $\eta = \rho\times v \times R$.  We find
%that the effect of these source of heating in the models explored is
%modest, with the mean \xco \ in our fiducial merger dropping by {$\sim
%  30 \%$}, and in the disc by less than a few \%.  Hereafter, we
%neglect this heating term, though note that including this term would
%only enhance the results presented in this paper.

The line cooling is assumed to occur via either CII or CO emission.
The fraction of hydrogen for which the carbon is mostly in the form of
CO is well-approximated by the following result from both
semi-analytic \citep{wol10} and numerical \citep{glo11} work:
\begin{equation}
\label{eq:abundance}
f_{\rm CO} = f_{\rm H2} \times e^{-4(0.53-0.045 {\rm
    ln}\frac{G_0'}{n_{\rm H}/{\rm cm^{-3}}}-0.097 {\rm ln}Z')/A_{\rm v}}
\end{equation}

When this fraction is above 50\%, we assume the cooling happens
predominantly via CO line cooling; else, the cooling occurs via CII
emission.  The cooling rate is calculated via an escape probability
formalism utilising the public code of \citet{kru07}.  We describe the
equations for the line radiative transfer (both within clouds, as is
pertinent to calculating the cooling rates, and across the model
galaxy, in \S~\ref{section:turtlebeach}).  

The dust cooling rate is:
\begin{equation}
\Lambda_{\rm dust} = \kappa (T_{\rm d})\mu_{\rm H}caT_{\rm d}^4. 
\end{equation}
We assume the bulk of the dust heating happens via IR radiation as IR
radiation likely dominates the heating over UV flux in the optically
thick centres of GMCs.  The IR radiation field is known from \sunrise
\ dust radiative transfer calculations (which will be described in
\S~\ref{section:sunrise}).

Finally, the dust and gas exchange energy via:
\begin{equation}
\Psi_{\rm gd} = \alpha_{\rm gd}n_{\rm H}T_{\rm g}^{1/2}(T_{\rm d}-T_{\rm g})
\end{equation}
where the thermal gas-dust exchange rate is $\alpha_{\rm gd} = 3.2
\times 10^{-34} Z'$ erg cm$^3$ K$^{-3/2}$ for \htwo, and $\alpha_{\rm
  gd } = 1 \times 10^{-33} Z'$ erg cm$^{3}$ K$^{-3/2}$ for HI
\citep{gol01}.

\subsection{Dust Radiative Transfer}
\label{section:sunrise}

In order to calculate the background radiation field from stars and the
dust temperature, we perform dust radiative transfer calculations
with the publicly available code \sunrise.  A full description of the
algorithms can be found in \citet{jon06a, jon10a}, and
\citet{jon10b}. Here, we summarise the aspects of the simulations most
relevant to this study.

The sources of radiation in the model galaxies are stellar clusters
and accreting black holes.  The stellar clusters emit a template
spectrum derived from \starburst \ calculations, with the
metallicities, masses and ages known from the \gadget \ simulations.
The AGN emits a spectrum based on observations of unreddened quasars
\citep{hop07}, though has little effect in the calculations here (see
the Appendix).

The substructure of the ISM on scales below the smoothing length of
the SPH simulations is unresolved.  We assume that star clusters with
ages $< 10$ Myr reside in natal birthclouds, and modulate their SED
accordingly.  These birthclouds contain HII regions and
photodissociation regions (PDRs) whose SEDs are calculated utilising
1D \mappings \ photoionisation models \citep{gro04,gro08,jon10a}.  The
time-averaged PDR covering fraction is a free-parameter.  We assume a
constant fraction of $f_{\rm PDR} = 0.3$, corresponding to a covering
lifetime of $\sim 2-3$ Myr \citep{gro08}. This value is motivated in
part by simulations by \citet{jon10a} which showed covering fractions
comparable to these result in synthetic SEDs of disc galaxies
comparable to the SINGS sample \citep{ken03}.  Changing this parameter
has minimal effects on the final results of this paper: we quantify
this and other potential effects of the subresolution modeling in
the Appendix.

When radiation leaves either the naked stellar cluster (with age $>
10$ Myr), or the HII region/PDR (for younger clusters), it is allowed
to interact with the diffuse ISM.  We assume the remaining cold
molecular phase has an negligible cross-section for interaction,
though test the effects of this assumption in the Appendix.

The dust mass in the diffuse ISM is calculated assuming a constant
dust to metals ratio of 0.4 \citep{dwe98,vla98,cal08}, where the
metallicity distribution is known from the SPH calculations. We use
the \citet{wei01} dust model with $R \equiv A_V/E_{B-V} = 3.15$, as
updated by \citet{dra07}. The dust and radiation field are assumed to
be in radiative equilibrium, utilising the methodology of
\citet{juv05} for calculating the converged radiation field. When the
radiation field has converged, we calculate the dust temperature in
each cell by iterating equations 6-8 of \citet{jon10b} utilising a
Newton-Raphson scheme.

\subsection{Molecular Line Radiative Transfer}
\label{section:turtlebeach}
Finally, with information about the spatial distribution of GMCs in
the model galaxies, and their mean \htwo \ fractions, densities,
temperatures, velocity dispersions and kinematics through the
galaxy, we are prepared to calculate the emergent CO line emission
from the model galaxy.  This involves two stages. First, we calculate
the escape probabilities of the CO lines from the GMCs.  We then track
the propagation of these photons through the model galaxy as they
potentially interact with other GMCs. 

Generally, CO line emission is set by the level populations.  The
source function from a given region for a given transition from
upper level to lower level $u\rightarrow l$ is given by:
\begin{equation}
S_{\nu}=\frac{n_{u}A_{ul}}{(n_{l}B_{lu}-n_{u}B_{ul})}
\end{equation}
where $A_{ul}, B_{lu}$ and $B_{ul}$ are the Einstein coefficients for
spontaneous emission, absorption, and stimulated emission,
respectively, and $n$ are the absolute level populations.

We first calculate the level populations within and probability for
photons to escape from the individual GMCs in the galaxy utilising the
publicly available code described in \citet{kru07}.  The levels are
assumed to be in statistical equilibrium and determined through the
rate equations:
\begin{eqnarray}
\label{eq:kmt_stateq}
\sum_l(C_{lu} + \beta_{lu}A_{lu})f_l =
\left[\sum_u(C_{ul}+\beta_{ul}A_{ul})\right]f_u\\
\sum_if_i = 1
\end{eqnarray}
where $C$ are the collisional rates, $f$ the fractional level
populations, and $\beta_{ul}$ is the escape probability for transition
$u\rightarrow l$.  The rate equations can be rearranged as an
eigenvalue problem, and solved accordingly.

The escape probability, $\beta_{ul}$ can be approximated by by
relating it to the optical depth in the line, $\tau_{ul}$
\citep{kru07}:
\begin{equation}
\label{eq:kmt_beta}
\beta_{ul} \approx \frac{1}{1+0.5\tau_{ul}}
\end{equation}

In the escape probability formalism, the optical depth of the line
through the cloud can be represented as:
\begin{equation}
\label{eq:kmt_tau}
\tau_{ul} =
\frac{g_u}{g_l}\frac{3A_{ul}\lambda_{ul}^3}{16(2\pi)^{3/2}\sigma}QN_{\rm
  H2}f_l\left(1-\frac{f_ug_l}{f_lg_u}\right)
\end{equation}
where $Q$ is the abundance of CO with respect to \htwo, $g_l$ and
$g_u$ are the statistical weights of the levels, $N_{\rm H2}$ is the
column density of \htwo \ through the cloud, $\lambda_{ul}$ is
the wavelength of the transition, and $\sigma$ is the velocity
dispersion in the cloud.  Equations
~\ref{eq:kmt_stateq}-\ref{eq:kmt_tau} are iterated upon utilising the
Newton-Raphson Method until the escape probabilities and level populations
within the GMCs are known.

With $\beta_{ul}$ calculated, we determine the effects of radiation
from individual GMCs on other GMCs in determining the final level
populations utilising the 3D non-LTE Monte Carlo radiative transfer
code \turtlebeach \ \citep{nar06b,nar08a}.  We begin with the level
populations found from the escape probability calculations as a guess,
and emit model photons from each GMC isotropically with direction
drawn randomly, and emission frequency drawn from a Gaussian profile
function:
\begin{equation}
\phi(\nu)=\frac{1}{\Delta \nu_{\rm D} \sqrt{\pi}}{\rm exp}\left \{-\left
(\nu-\nu_0-\bf{v \cdot \hat n} \frac{\nu_{ul}}{c}\right )^2 /
\Delta \nu_{\rm D}^2\right \}
\end{equation}
where $\nu_0$ is the rest frequency of the line, $v$ is the velocity
of the cloud in the direction of the photon's emission, $c$ is the
speed of light, and $\Delta \nu_{\rm D}$ is the doppler-width of the
emission line.

When the photon passes through a cell, it interacts with a GMC and
sees an opacity of:
\begin{equation}
\alpha_\nu^{ul}(\rm gas)= V_{\rm fill}\left[\frac{h
    \nu_{ul}}{4\pi}\phi(\nu)(\it n_lB_{lu}-n_uB_{ul})\right]
\end{equation}
where ${\rm V}_{\rm fill}$ is the volume filling factor of the spherical
GMC.  We neglect absorption by dust in this model.

After all GMCs have emitted some number of model photons, the level
populations in the GMCs are updated by assuming detailed balance:
\begin{eqnarray}
\label{eq:tb_stateq}
  n_l\left [\sum_{k<l}\beta_{lk}A_{lk}+\sum_{k\neq
      l}(B_{lk}J_{\nu}+C_{lk})\right]=
  \\\nonumber \sum_{k>l}n_k\beta_{kl}A_{kl}+\sum_{k\neq
    l}n_k(B_{kl}J_\nu+C_{kl})
\end{eqnarray}
where $C_{lk}$ and $C_{kl}$ are the collisional rates, and $\beta$
only exists for transition $k\rightarrow l$ such that $k=l+1$.
Equations~\ref{eq:tb_stateq} are solved via Gauss-Jordan matrix
inversion.

This process is iterated upon until the level populations have
achieved convergence.  Here, we demand that they not vary by more than
a fractional difference of $1\times10^{-3}$ for at least 3 iterations.

Once the level populations have been solved for, we build the formal
spectrum by choosing an (arbitrary) viewing angle, and integrating
along lines of sight \citep[e.g.][]{wal94b}: 
\begin{equation}
  I_\nu = \sum_{z_0}^{z}S_\nu (z) \left [ 1-e^{-\tau_\nu(z)} \right
  ]e^{-\tau_\nu(\rm tot)}
\end{equation}

Tests of \turtlebeach \ against the publicly available Leiden
Benchmarks \citep{van02} are presented in \citet{nar06b}.  We obtained
our coefficients from the {\it Leiden Atomic and Molecular Database}
\citep{sch05}.  We assume a fractional carbon abundance of
$1.5\times10^{-4}$, though the abundance of CO with respect to \htwo
\ is given by Equation~\ref{eq:abundance}.

\begin{figure*}
\vspace{-1cm}
\includegraphics[scale=0.275]{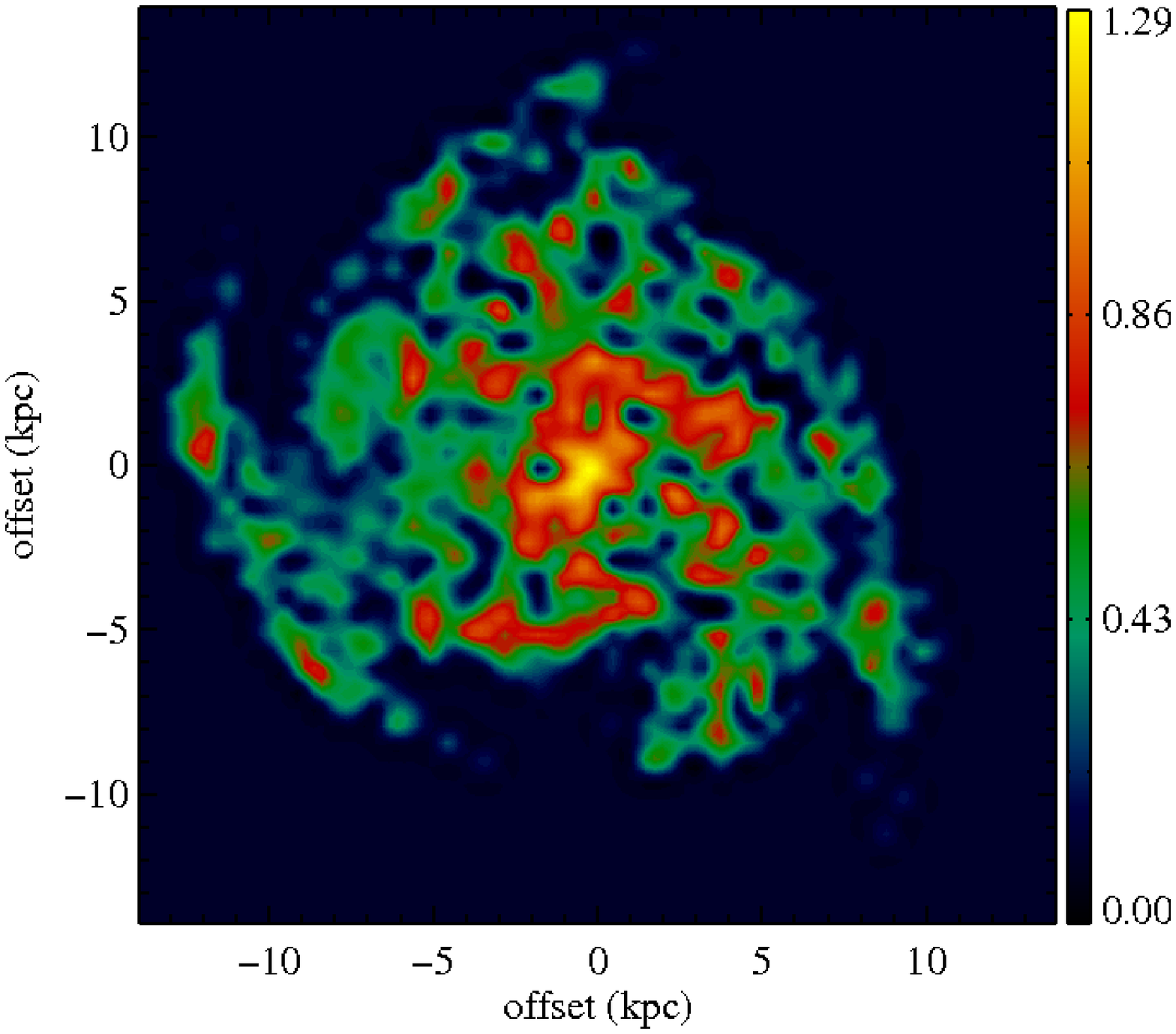}
\includegraphics[scale=0.275]{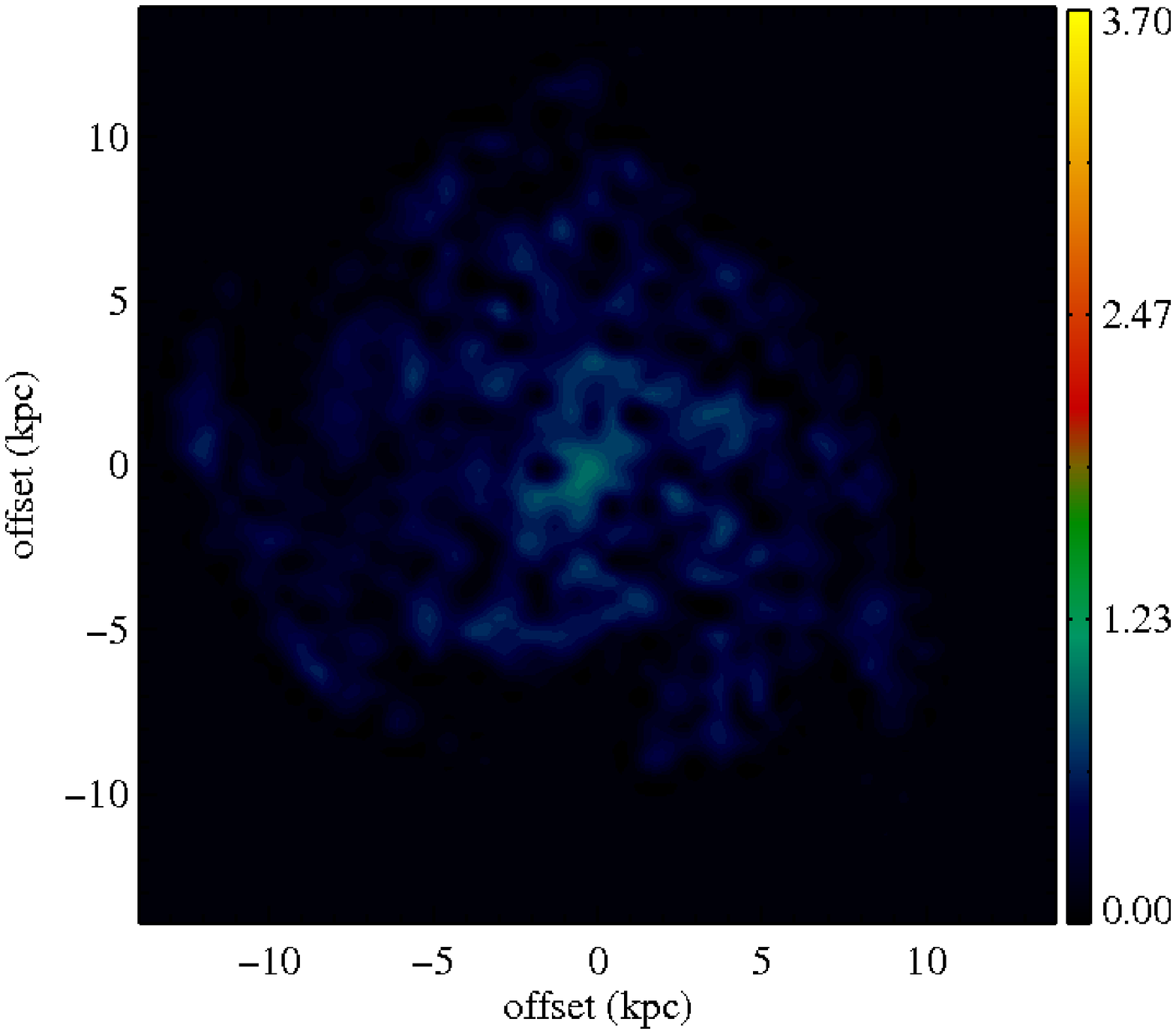}
\includegraphics[scale=0.275]{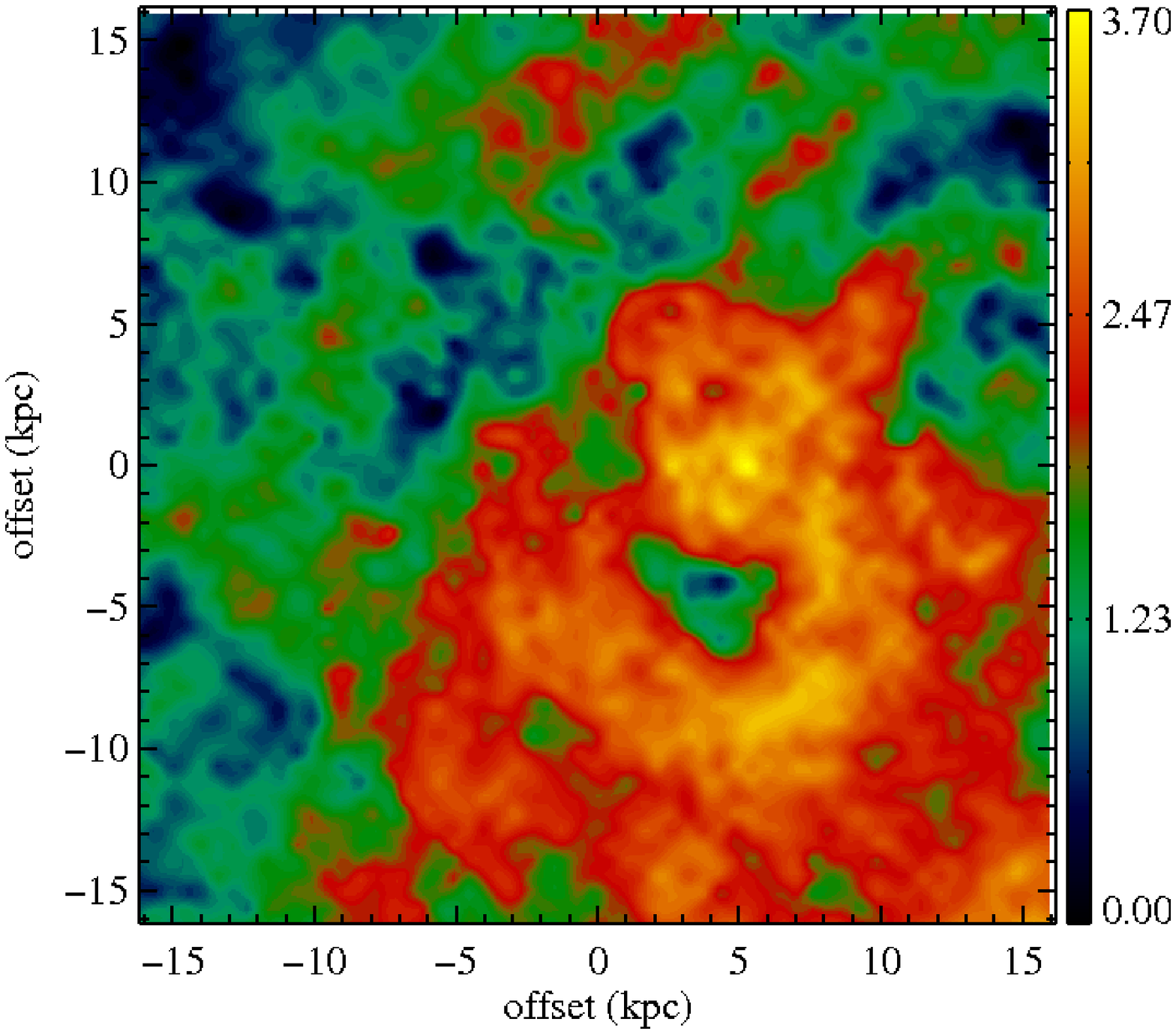}
\vspace{-1cm}
\caption{Velocity integrated brightness temperature $W_{\rm CO}$ maps
  for model disc galaxy (left), and fiducial merger snapshot (right).
  The colour scales correspond to the colour bars on the right of each
  panel, and the units are log$_{\rm 10}$ (K-\kmsend). The centre
  panel shows the model disc again, but with the same colour scale as
  the merger for comparison.  Low gas kinetic temperature ($\sim10$ K)
  and velocity dispersions cause the bulk of the disc galaxy to have
  intensities of $\sim10$ K-\kmsend.  In contrast, the velocity
  dispersion within GMCs in the ULIRG can be many tens of \kmsend,
  with gas kinetic temperatures near 50 K.  Summed over a sightline,
  the observed gas intensity can be $>10^3
  $K-\kmsend. \label{figure:iso_d3_map}}
\end{figure*}

\section{Observational and Physical Properties of Simulated Galaxies}
\label{section:observationalproperties}

As we aim to compare potential variations in \xco \ in our simulated
galaxy mergers to those that are actually observed, it is worth
briefly comparing the physical and synthetic observational properties
of our model galaxies to real galaxies.  Our fiducial merger has been
well-studied in the literature, and is very much an average merger
simulation as far as the range of simulated SFRs, black hole accretion
rates and bolometric luminosities.  While the processes described in
this section generically describe gas-rich mergers, what we
summarise here has been calculated and published previously explicitly
for our fiducial model.

 The merger goes through elevated star formation rate upon first
 passage as tidal torques on the gas cause the gas to lose angular
 momentum and fall toward the centres causing high-density regions
 \citep{mih96,hop06,nar08b,jun09}.  The galaxy undergoes a starburst
 upon final coalescence, and peaks in its bolometric luminosity.
 Radiative transfer post-processing on these models have found that
 the model galaxy is then visible as a ULIRG \citep{cha07}. The same
 gaseous inflows can drive sufficient black hole growth to result in
 optical quasar activity \citep{dim05,hop05a,hop05b,hop06}.  Effects
 of the AGN feedback can be seen in both the warm infrared colours of
 the galaxy \citep{you09}, as well as molecular outflows
 \citep{nar06a,nar08a}.

The truncation of the starburst by a combination of gas consumption
and AGN feedback can render the galaxy observable as an E+A
post-starburst \citep{sny11} before it evolves into into a dead
early-type \citep{spr05c,hop07c} with colours comparable to those
observed on the red-sequence \citep{spr05c,hop08a,hop08b}.  The
kinematic \citep{cox06b}, X-ray \citep{cox06c}, nuclear emission
\citep{hop08c,hop09b}, and molecular disc properties \citep{xu10} of
this merger remnant have all been studied and found to be comparable
to those observed. Similarly, the remnant lies on the fundamental
plane \citep{dim05,rob06a,hop08e}.

\section{Results}
\label{section:results}

\subsection{GMCs in ``Normal'' Discs}
\label{section:xco_dics}

\begin{figure*}
\hspace{-1cm}
\includegraphics[angle=90,scale=0.8]{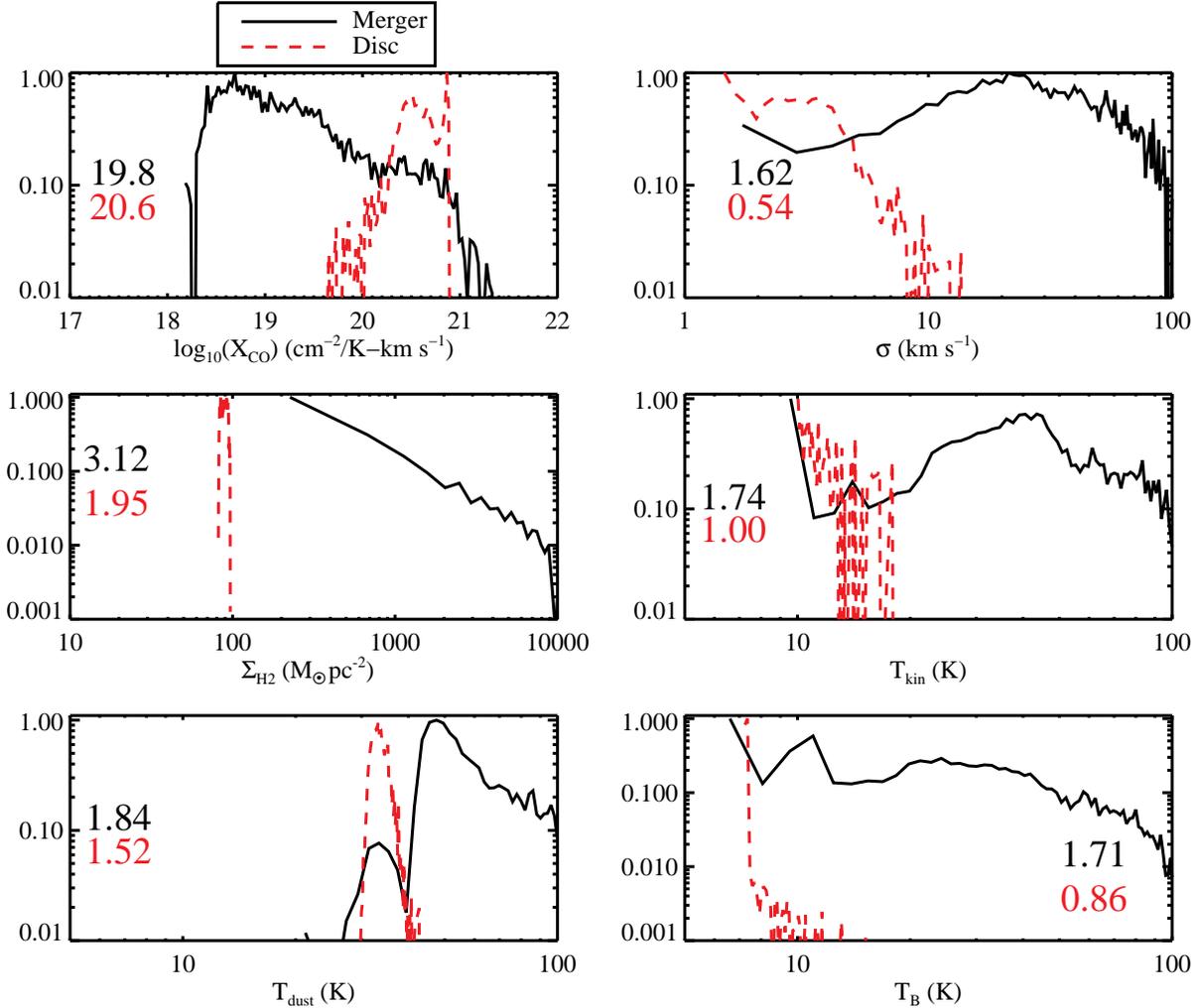}
\caption{Emission-weighted distributions of \xco \ values and physical
  properties for the GMCs in the model disc galaxy (red-dashed line)
  and fiducial merger model (solid-black line).  The ordinate values
  are normalised.  Starting from the top left, and going clock-wise,
  the plots show \xco, velocity dispersion, kinetic temperature,
  brightness temperature, dust temperature, and GMC surface
  density. The disc galaxy predominantly has GMCs with physical
  properties comparable to the Milky Way's, and thus has similar \xco
  \ values.  The GMCs in the merger show a broad distribution in \xco
  \ values, with a lower mean than the disc galaxy.  The lower \xco
  \ owes to larger gas temperatures (which are larger due to efficient
  coupling with the warm dust at high densities) and large velocity
  dispersions in the gas.  The numbers in each panel refer to the
  log$_{10}$ of the emission-weighted mean value, and the black (top)
  number corresponds to the merger whilst the red (bottom) corresponds
  to the disc. Because the numbers correspond to the log$_{10}$ of the
  mean in the physical quantities, they will have larger values than
  one would pick by eye in the log-log
  plots.\label{figure:xco_distribution} }
\end{figure*}

In the far left panel of Figure~\ref{figure:iso_d3_map}, we show the
velocity-integrated brightness temperature map of the model disc
galaxy.  As expected, the central regions are the brightest, and the
outer disc has little CO emission.  In the top left panel of
Figure~\ref{figure:xco_distribution}, we plot the emission-weighted
distribution of \xco \ values for the GMCs in our model disc galaxy
and the fiducial model merger.  We additionally plot the distribution
of GMC physical properties in both the disc and merger.  We will
return to this plot frequently throughout this section and the next.

 The luminosity-weighted \xco \ in our model disc is $\sim 4 \times
 10^{20}$ \ \xcounits \ with a relatively narrow dispersion.  The
 dispersion is narrow because the surface densities, kinetic
 temperatures and velocity dispersions of the model disc GMCs show
 fairly little variation.  To remind the reader, the column densities
 in the GMCs in our disc galaxy are set to be the surface density of
 cold gas in the cell.  When the GMC is unresolved in the simulation,
 we set the subgrid value of the surface density to $\Sigma_{\rm
   cloud} = 100 \ \msun/{\rm pc}^{2}$.  This value was chosen to match
 the roughly constant surface density of Galactic molecular clouds.
 Nearly all of the GMCs in the model disc take on this value for a
 surface density.

The kinetic temperatures of GMCs in the disc have a relatively tight
distribution near 10 K, as shown in
Figure~\ref{figure:xco_distribution}. Because the GMCs have a
relatively low density compared with starbursts (the mass-weighted
value is $\sim500$ \cmthree), there is little coupling with the dust
grains (which are a factor of a few hotter;
Figure~\ref{figure:xco_distribution}).  Thus the temperature is
primarily determined by molecular line cooling, and heating by cosmic
rays and the grain photoelectric effect.  The kinetic temperature
helps to set the brightness temperature, though the two are not
identical.  The emission-weighted brightness temperature for the
merger(disc) are $\sim 50$(7)K.

Finally, the distribution of velocity dispersions in the GMCs is
fairly narrow.  Recalling \S~\ref{section:methods}, the velocity
dispersion of the clouds is taken by calculating the dispersion
amongst the cell's nearest neighbours, with a subgrid model for
unresolved clouds (Equation~\ref{equation:sigma}).  Because the disc
is dynamically cold, the velocity dispersions are primarily set by the
latter case.  This results in an emission-weighted velocity dispersion
within GMCs in the model disc of $\sim 3$ \kmsend, with a maximum of
$\sim15$ \kmsend.  These values compare favourably with the velocity
dispersions reported in the comprehensive survey of \citet{sol87}, and
the more recent review by \citet{bli07}.

We can ask why the simulated \xco \ from the model galaxy is
comparable to the Galactic average, $\xco \approx 2-4 \times 10^{20}
\xcounits.$ In principle this occurs because the physical conditions
in the model GMCs by and large match those of observed GMCs in the
Milky Way.  In this sense, the fact that our model value for \xco \ in
quiescent discs matches that of the Galaxy is by construction.
However, there are two salient points here.

First, it is important to remember that we allow for the possibility
that the galactic environment can set the physical conditions in the
GMCs if the pressure is sufficiently high.  The fact that the default
value for the surface density and velocity dispersions in the clouds
is typically used is a statement that the galactic environment in the
model disc galaxy is not sufficiently extreme to cause significant
changes in the surface densities, temperatures, or velocity
dispersions in the GMCs from the Galactic values.  As we will see in
the subsequent section, this is not the case in mergers.

Second, the subresolution values for the GMCs are not without
physics. GMCs through the Local Group are observed to obey the
\citet{lar81} relations: they follow a linewidth-size relationship
with $\sigma \propto R^{0.5}$, they have virial parameters $\alpha\sim
1$, and they all have roughly the same surface density $\Sigma\sim
100$ $\msun$ pc$^{-2}$ \citep{bli07,bol08}. The origin of these
observed relationships is debated, but their universality argues for
some sort of internal regulation mechanism operating in GMCs
\citep[e.g. ][]{kru06,she08}. Regardless of the underlying mechanism,
though, our subgrid model is not simply tuned to reproduce the
``right" $X_{\rm CO}$. Instead, it models the real physical properties
of GMCs\footnote{We note, however, that clouds need not be virialised
  to have \xco \ comparable to observed galactic values.  Provided
  that $\Sigma_{\rm cloud}$, $T$, and $\sigma$ remain within a modest
  range of values, $\xco \sim 2-4 \times 10^{20} \xcounits$
  \citep{she11b}.}.

Finally, it is important to note that the results presented in this
section do not necessarily translate to disc galaxies at
high-redshift.  Galaxies on the ``main-sequence'' of star formation
rates at high-\z \ \citep[e.g. ][]{noe07b,noe07a} still form stars at
rates comparable to present-epoch mergers \citep{dad05,dad07} because
they have very high gaseous surface densities, though they may be
morphologically classified as discs \citep{for09}.

\subsection{\xco \ in Merger-Driven Starbursts}
\label{section:xco_starbursts}

We now turn to \xco \ in galaxy mergers.  Before embarking on the
remainder of this section, it is important to emphasise that the many
of the GMCs in the model starburst are resolved (this is clear from
the $\Sigma_{\rm H2}$ panel in Figure~\ref{figure:xco_distribution}).
Thus, the derived values for \xco \ are independent of subresolution
assumptions.

During the merger, gas is funneled toward the nuclear regions, causing
dense concentrations of molecular gas \citep{bar91,bar96}.  The
surface densities of the GMCs in our simulations rise accordingly.  In
principle, this would cause a rise in the CO-\htwo \ conversion factor
(c.f. Equation~\ref{equation:xco_units}).  However, during the
merger-induced starburst, the increase in velocity-integrated line
intensity exceeds the rise in surface density, causing \xco \ to drop
from the Galactic value.

\begin{figure*}
\hspace{-1cm}
\includegraphics[scale=0.55]{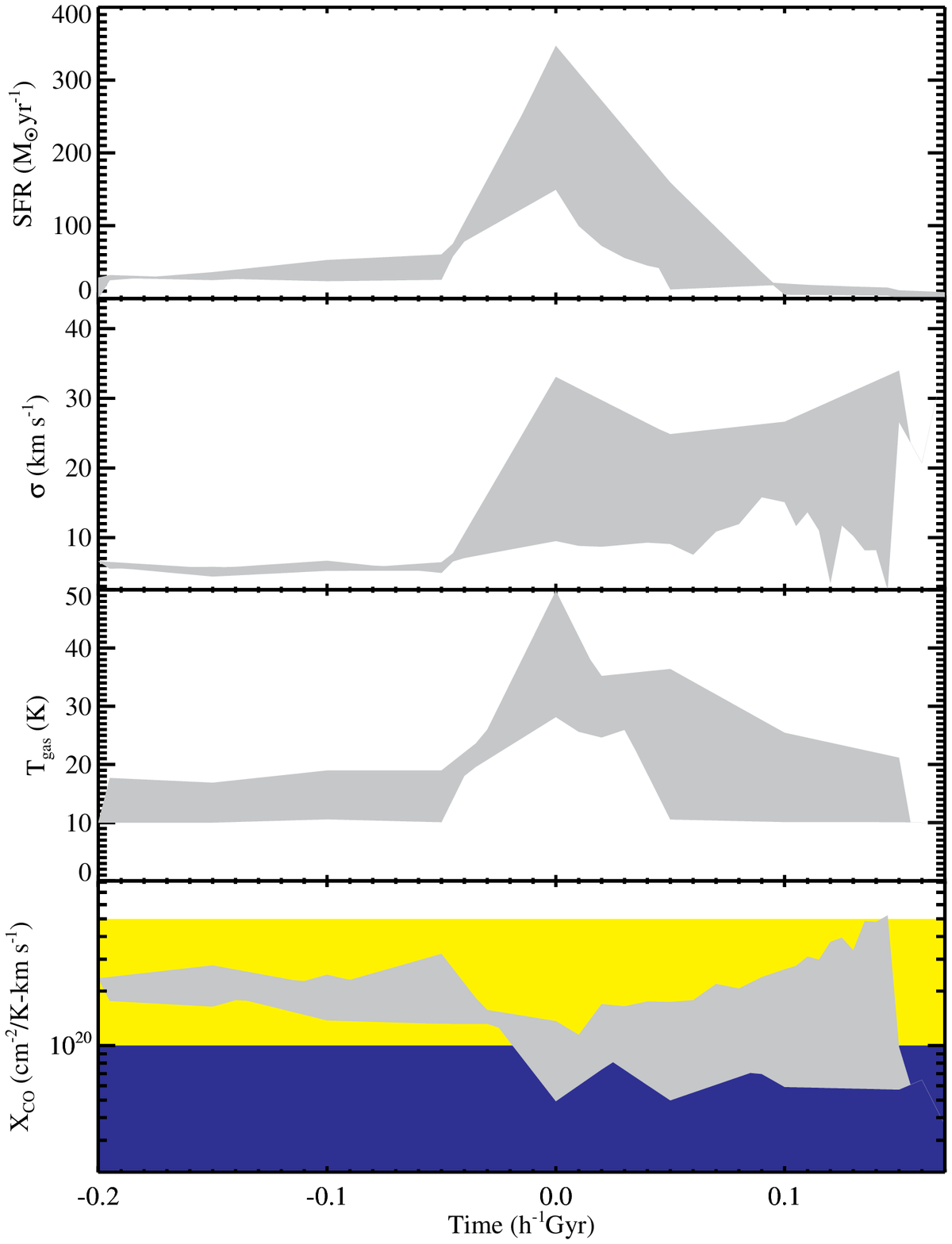}
\caption{The evolution of the star formation rate, emission-weighted
  velocity dispersion, kinetic temperature, and \xco \ of the
  individual GMCs within the galaxies for all three merger models as a
  function of time. The time axes are centred for each model about the
  point of maximum star formation rate.  The grey shaded region
  denotes the range in emission-weighted mean values for all three
  models. This means that at a given time step, the shaded region is
  defined by the maximum and minimum value of a given quantity seen
  among the three merger models.  In the bottom panel, the yellow and
  blue bands denote the typical ranges of \xco \ observed for the
  Galaxy and ULIRGs, respectively \citep[as compiled by][]{tac08}.
  Prior to the burst, the inspiralling discs have \xco \ values
  comparable to the Galactic mean.  Upon the merger, increased
  velocity dispersions and gas temperatures contribute to lowering
  \xco.  In the post-merger stage, differences in \htwo \ abundances,
  CO abundances, and time for the gas to re-virialise contribute
  toward a large dispersion in \xco
  \ values. \label{figure:xco_evolution}}
\end{figure*}

In Figure~\ref{figure:xco_evolution}, we show the evolution of the
star formation rate, gas temperature, velocity dispersion and \xco\ as
a function of time for the three model galaxy mergers.  The shaded
region denotes the range of mean values among the GMCs within the
merger models at each timestep, i.e. at time 0 the lowest point
outlined in gray corresponds to the lowest galaxy-averaged value of
the three merger models, and the highest point in gray corresponds to
the highest galaxy-averaged value among the three.  The time axes are
centred around the peak in the starburst for each model.  When the
galaxies merge, the discs are destroyed.  During this time, the
dominant contributor to $\sigma$ within the GMCs is the nonthermal
component derived from the local resolved velocity dispersion of the
gas.  The nonthermal velocity dispersion is driven by the dynamics
during the galaxy merger and mixing of stellar mass with the \htwo
\ gas.

During final coalescence in the merger, when the SFR peaks at a few
hundred \msunyrend, the fraction of dense gas rises, a result verified
both in theoretical models
\citep[e.g. ][]{mih94a,mih96,nar08d,nar08b,bou11}, and observations
\citep{jun09}.  The mass-weighted mean GMC density rises to roughly
$\ga10^{4}$ \cmthree, compared to $\sim500 $ \cmthree \ in the model
disc.  At these high densities, the energy exchange between dust and
gas becomes efficient, and the gas temperatures begin to approach the
dust temperatures.  At the same time, the dust is being heated by an
amplified radiation field due to the merger-induced starburst.  This
is demonstrated explicitly in Figure~\ref{figure:xco_distribution},
where we show the dust and gas (kinetic) temperature distributions of
the GMCs in the model galaxies.  The mean gas temperature is higher by
a factor of a few than the roughly $\sim10$ K GMCs in the model disc.
The rise in the gas kinetic temperature during the starburst is shown
in Figure~\ref{figure:xco_evolution} as well\footnote{We remind the
  reader that we adopt a constant cosmic ray ionisation rate in all
  models.  If cosmic ray energy densities increase in starburst
  environments as suggested by recent observational \citep{abd10b} and
  theoretical \citep{pap10a,pap10b} work, then the gas temperatures
  would further increase, causing \xco \ in starbursts to fall even
  further \citep[though see ][for an expanded study on the role of
    cosmic rays in the $X$-factor at various extinctions]{bel06}.}.

The large molecular gas densities in the merger also mean the CO is
thermalised in the ground state transition.  When level populations
are in local thermodynamic equilibrium (LTE), their source function
can be described by the Planck Function. In this limit, the source
function rises with temperature.  Hence, the rise in gas kinetic
temperature during the burst contributes to driving \xco \ down.

The combination of the increased velocity dispersion and the
brightness temperature combine to exceed the increase in surface
density, which causes a depressed mean \xco \ from the Galactic value
during the merger.  We now return to the first panel
Figure~\ref{figure:xco_distribution} to explicitly compare \xco \ in
the merger against the disc galaxy.  We see that \xco \ has a broad
distribution for the model merger.  A number of GMCs outside of the
nucleus are similar to the disc galaxy's in terms of their physical
properties.  These GMCs are unresolved in our simulations (owing to
the fact that they reside in lower-density environments), and thus
take on surface density and velocity dispersion values comparable to
those observed in the Galaxy.  However, the GMCs toward the centre of
the galaxy all have larger surface densities, velocity dispersions,
and kinetic temperatures.  The latter two combine to depress \xco
\ compared to the values seen in the disc by a factor of $\sim 5-10$.
Because most of the mass in the merger is in the central regions, the
luminosity-weighted mean is low.  Test simulations with fixed
temperatures or velocity dispersions show that the increased kinetic
temperature and velocity dispersion in the gas contribute roughly
equally to the increased line intensity in the merger simulation.
This is somewhat apparent from Figure~\ref{figure:xco_distribution},
where we see similar distributions values for the kinetic temperatures
and velocity dispersions in the gas.

\begin{figure*}
\includegraphics[scale=0.3]{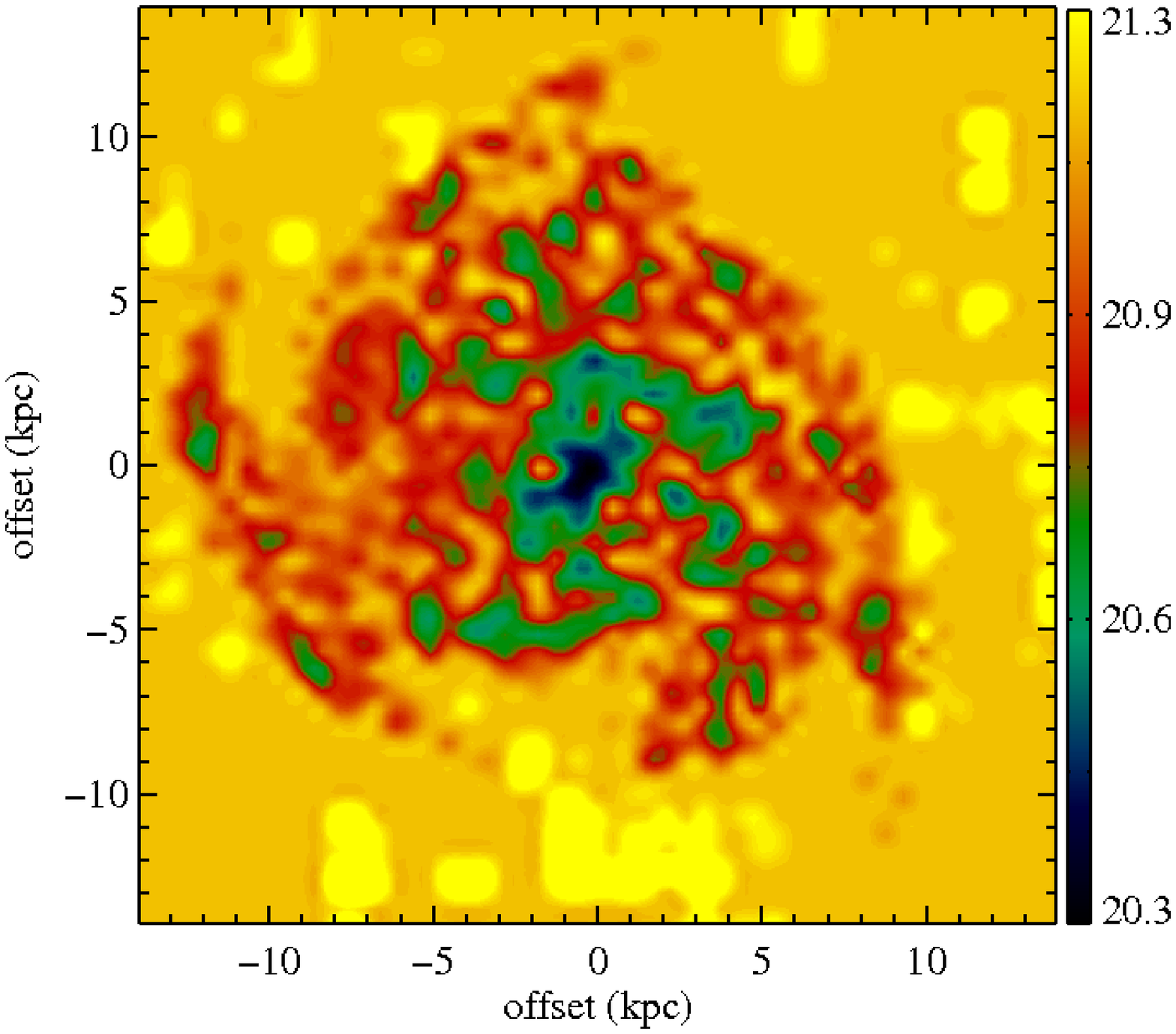}
\includegraphics[scale=0.3]{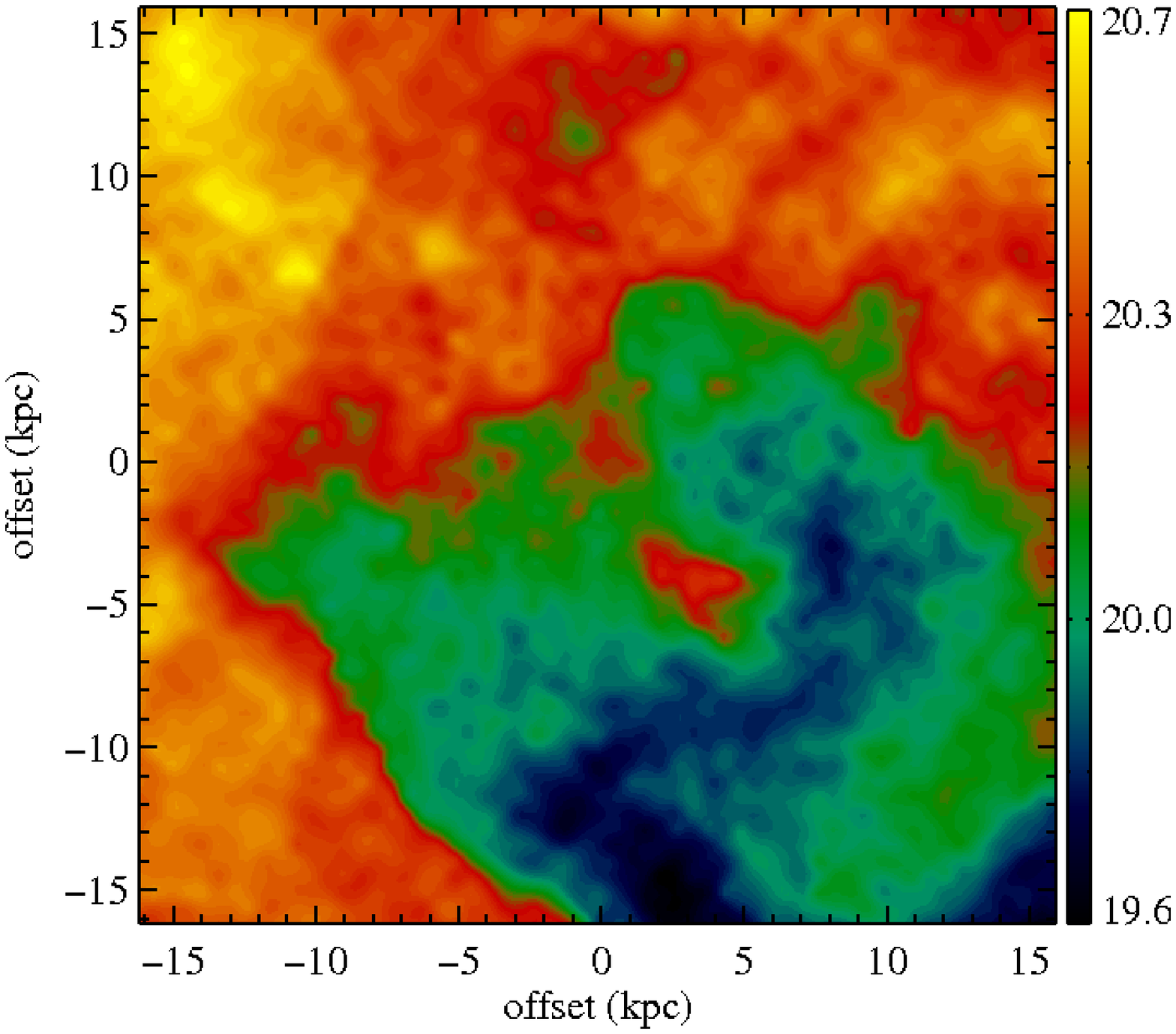}
\caption{Maps of \xco \ for the fiducial disc galaxy and merger. The
  colour scale denotes log$_{\rm 10}$ (\xcounits). \xco \ is lower
  toward the centre of the disc galaxy due to higher temperatures and
  velocity dispersions in the clouds.  The warm and high-$\sigma$ gas
  is somewhat more spread out in the merger.  We enforce a maximum
  \xco \ in the color bar of the disc galaxy of 21.3 to aid in
  clarity, though there are a few pixels with values as high as
  21.8.\label{figure:xco_map}}

\end{figure*}

The magnitude by which \xco \ decreases is dependent on the strength
of the merger.  Turning to Figure~\ref{figure:xco_evolution}, we see
a range in \xco \ values during the burst.  The model with the
largest \xco \ during the burst corresponds to the lowest peak SFR.  A
key point of this aspect of the model is that there is no ``merger
value'' of \xco: \xco \ depends on the physical parameters of the
emitting galaxy.

What happens in the post-starburst stage is also highly
merger-specific.  During this phase, the galaxy is a gas-poor early
type.  Generally, the gas has a large velocity dispersion for at least
a dynamical time after the burst.  This is consistent with what was
seen in simulations of CO gas in high-\z \ submillimetre galaxies
\citep{nar09}.  During this phase, it is less trivial to simply relate
the observed \xco \ to the gas velocity dispersion and temperature as
there is a much larger dispersion in molecular gas fractions and CO
abundances.  This owes to the fact that there are highly varying
physical conditions in the post-burst galaxy, which drive strong
variations in the \htwo \ and CO abundances. Some of the simulations
return to a Galactic \xco \ value quickly, while others remain low.

In summary, during the merger-induced starburst, \xco \ drops in
galaxies from the standard Galactic value due to increased gas
temperatures and velocity dispersions.  During this time, the CO
abundances are $\sim 1\times10^{-4}$/\htwo \ and molecular gas fractions
near unity in the main CO emitting region.  In the resulting gas-poor
merger remnant, the dynamical and thermal history can vary from model
to model, and the evolution of \xco \ is less uniform among mergers.

\subsection{The Variation of \xco \ with Galactocentric Radius}
\label{section:location}

\begin{figure}
\includegraphics[scale=0.5]{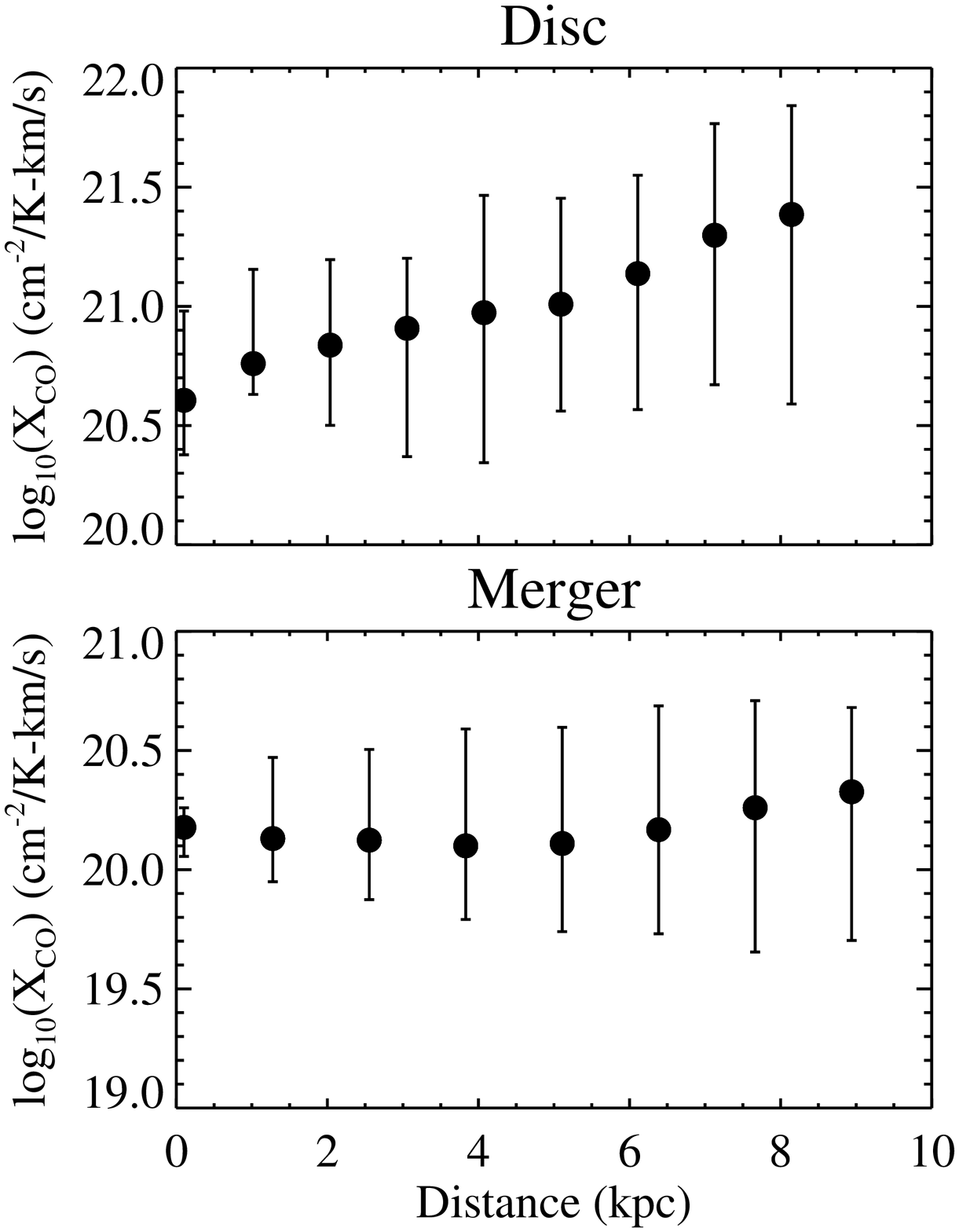}
\caption{\xco \ as a function of spatial distribution for the
  molecular gas in model disc (top) and merger (bottom).  The \xco
  \ is derived from the maps (Figure~\ref{figure:xco_map}), and is the
  emission-weighted mean \xco \ in bins of galactocentric distance.
  The bars around the points represent the range of \xco \ values
  within a given distance bin.  The molecular gas at the centre of the
  disc has systematically lower \xco \ values than the outer disc.  In the
  merger, \xco \ shows a wide-range of values throughout the galaxy.
 \label{figure:xco_spatial_distribution}}
\end{figure}

With the concepts presented in \S~\ref{section:xco_starbursts}, we are
now in a position to understand how \xco \ varies in galaxies as a
function of spatial location.  In Figure~\ref{figure:xco_map}, we show
the simulated \xco \ maps for the disc galaxy and fiducial merger, and
in Figure~\ref{figure:xco_spatial_distribution}, we plot the values
for \xco \ in the GMCs in our model disc galaxy and fiducial merger as
a function of radius from the centre of the galaxy.  The \xco \ values
from Figure~\ref{figure:xco_spatial_distribution} come from the map in
Figure~\ref{figure:xco_map}.  The \xco \ values are binned in bins of
distance, and represent the emission-weighted mean within a given
distance bin.  The bars denote the range of \xco \ values seen in a
given distance bin.

\xco \ in the centre of the model disc galaxy is systematically lower
than in the rest of the galaxy.  In particular, a number of GMCs along
the line of sight have velocity dispersions larger than the typical
virialised values, with values elevated by a factor of $\sim 2$.
Similarly, due to the elevated densities in the nucleus combined with
a warmer dust temperature, the gas temperatures of some GMCs can reach
values up to 15 K.  This causes \xco \ in the central kiloparsec to
generally display the lowest values in the galaxy.  Depressed values
of \xco \ from the Galactic mean have been observed in at least a few
GMCs toward the Galactic Centre \citep{oka98}.  It is important to
note that the regions where \xco $> 10^{21} \xcounits$ represents much
of the area, but a negligible fraction of the gas mass in the galaxy.
This is evident from Figure~\ref{figure:xco_distribution}.

In the fiducial model merger, unlike the situation with the model disc
galaxy, we see no clear trend in \xco \ with galactocentric radius.
Due to the violent nature of the gasdynamics during the merger, gas of
a variety of physical conditions is mixed together.  Consequently, we
see a large range of $X$-factors in the GMCs throughout the galaxy.

Because the emission from the merger is irregular, it is possible that
by choosing a different centre, the results from
Figure~\ref{figure:xco_spatial_distribution} would change.  To test
this, we recentered the image on the peak of the velocity-integrated
intensity.  Doing this provides no substantial change in the results
of Figure~\ref{figure:xco_spatial_distribution}.

\section{Discussion}
\label{section:discussion}

\subsection{Observational Consequences of the Model}

We have presented a model in which \xco \ in GMCs is dependent on the
physical conditions within the clouds.  When the surface densities,
kinetic temperatures and velocity dispersions within the GMCs resemble
those of observed clouds in the Galaxy, the resulting \xco \ factor is
comparable to the observed Galactic mean value.  In starbursts, while
the surface densities of clouds are higher, this is offset by both
larger velocity dispersions in the GMCs as well as larger gas
temperatures.  The increased linewidths represent the turbulent
velocity dispersion in the merger, as well as the stellar potential.
The increased gas temperatures owe to efficient coupling with the dust
at the high densities encountered in a merger.  A fundamental point of
this study is that the physical conditions which cause \xco \ to vary
in starbursts are coupled.  The same processes which drive the
increased gas surface density also cause an increase in star formation
rate which drives up the dust and consequently the gas temperatures.
Similarly, in a merger-driven burst, the gas velocity dispersion rises
during the merger. 

While it is of utmost importance to parameterise \xco \ in terms of
observable properties, because the physical parameters which drive
observed values of \xco \ are coupled, this is a nontrivial task which
is outside the scope of this work (though it will be investigated in a
forthcoming paper).  Empirically, there is a tentative trend that \xco
\ decreases with increasing galaxy surface density \citep[][]{tac08}.
In the context of the models presented here, such a trend is plausible
\citep[see also][]{she11b}.  One might expect that higher surface
density systems typically arise in situations when the velocity
dispersion is high and the star formation rates, dust temperatures and
gas temperatures are also high.

While the models investigated here by no means comprise an exhaustive
parameter-space study of galaxy masses, merger mass ratios, or merger
orbits, we can investigate whether \xco \ can be parameterised by
$\Sigma_{\rm H2}$ in the simulations.  To increase the dynamic range
of surface densities in our models, we include one additional
simulation of a high-redshift merger.  The merger is the model
submillimetre galaxy of \citet{hay11} during the coalescence when the
peak merger-induced starburst is $\sim 4500$ \msunyrend. The model
submillimetre galaxy has been shown to reproduce both the observed SED
\citep{nar10a}, CO properties \citep{nar09}, overlap with 24 \micron
\ sources \citep{nar10b}, and number counts of observed SMGs
\citep{hay10}.  Similarly, to increase the number of galaxies in our
sample, we include many snapshots for the mergers (i.e. not just the
snapshots at peak SFR).

\begin{figure}
\hspace{-1cm}
\includegraphics[scale=0.4,angle=90]{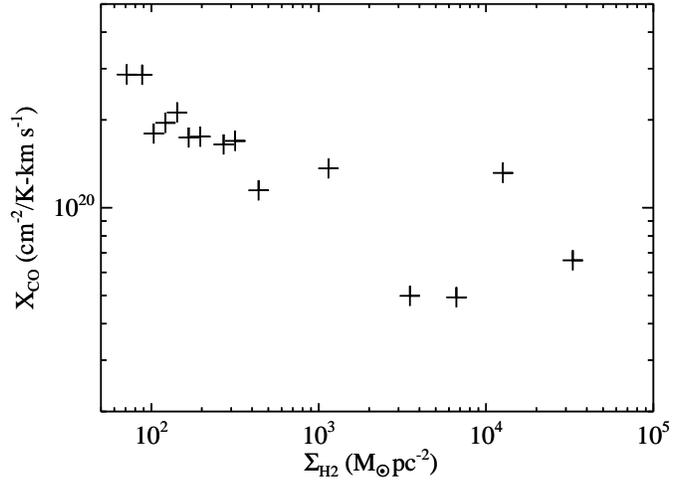}
\caption{Emission-weighted mean \xco \ in GMCs for each model galaxy
  versus their emission-weighted mean surface density.  The points
  represent the mean values for all of the GMCs within individual
  galaxies, and the galaxies are binned by surface density. In order
  to increase our sample size, many snapshots (i.e. not just the peak
  SFR for the mergers), as well as a model submillimetre galaxy, were
  included in this plot. Higher surface density environments tend to
  correspond with merger-driven starbursts in our models, and hence
  larger values of $\sigma$ and $T_{\rm K}$.  In these cases, \xco
  \ tends to be lower than the standard Galactic
  value.  \label{figure:xco_surfacedensity}}
\end{figure}

We plot the emission-weighted mean \xco \ of the GMCs in our model
galaxies versus their emission-weighted mean surface densities in
Figure~\ref{figure:xco_surfacedensity}.  The models include the model
mergers described in \S~\ref{section:methods}, the model disc galaxy,
and the newly introduced model submillimetre galaxy.  The galaxies are
binned by $\Sigma_{\rm H2}$.  The trend seen is what is expected: that
\xco \ should decrease with gas surface density.  In principle this
owes to the fact that the highest surface density galaxies in our
simulations are also forming stars at $10^2-10^3$ \msunyrend, and have
relatively large velocity dispersions.  However, we emphasise strongly
that Figure~\ref{figure:xco_surfacedensity} is to be taken as a
qualitative trend, rather than robust.  A larger parameter-space
survey of the simulations will be undertaken for a future study to
identify the true mean and dispersions of \xco \ as a function of
$\Sigma_{\rm H2}$.

The variation in \xco \ with environment may have implications for
observed Kennicutt-Schmidt star formation laws.  An example of this
was presented by \citet{dad10b} and \citet{gen10} who applied a
starburst \xco \ value to the inferred mergers in their observed
sample of galaxies, and a Galactic \xco \ value to the discs.  Doing
so results in a bimodal star formation rate surface density-gas
surface density relation. On the other hand, \citet{ost11} pointed out
that if a simple $\Sigma$-dependent \xco $\propto \Sigma^{-0.5}$ is used
above $\Sigma_{\rm H2}=100$ Msun/pc$^2$, a unimodal empirical star
formation relation results (with $\Sigma_{\rm SFR} \propto \Sigma^2$,
consistent with theoretical expectations for self-regulated star
formation in this regime).  

 Interpreting results for high-redshift galaxies in the context of our
 model is complex. Our model advocates for lower \xco \ values in high
 surface density environments.  However, galaxies at \zsim 2 which
 have surface densities comparable to local ULIRGs may in fact be
 discs \citep[e.g. ][]{dad05,gen06,for09,hop10}.  How \xco \ of
 high-\z \ discs should scale in this model is unclear at present.
 While their surface densities are comparable to local ULIRGs, their
 velocity dispersions may not show the same enhancement as seen in our
 model mergers.  However, their gas temperatures may be comparable to
 their dust temperatures if the densities are high enough.  Work by
 the Chicago group (R. Feldmann et al. in prep.) is underway to
 investigate this.  Either way, the fact that our results tentatively
 suggest a relationship between \xco \ and surface density implies a
 continuum in \xco \ values, rather than a bimodality.  Thus the
 relationship between our model results and the interpretation of
 high-\z \ observations by \citet{dad10b} and \citet{gen10} will
 depend on the distribution of surface densities in their observed
 galaxies, among other issues.  We note, however, that our work, like
 that of \citet{tey10}, is consistent with the idea that the observed
 behaviour in the star formation rates and \xco \ can be explained
 without the need to invoke a volumetric star formation law that is
 different in discs and mergers.  The change in \xco \ we see in our
 simulations occurs because of changes in the physical conditions of
 GMCs associated with the merger, and not because the underlying star
 formation law is different.

Finally, the concepts presented in this paper are testable in the near
future with ALMA.  Our models suggest that high spatial resolution
observations of nearby ULIRGs will display both large velocity
dispersions in the CO gas, and larger brightness temperatures
than those seen in observations of Galactic GMCs on a comparable
scale.  We see this when comparing the panels of
Figure~\ref{figure:iso_d3_map}. Some observational evidence for this
already exists. Interferometric surveys of the central regions of
nearby ULIRGs show velocity dispersions of hundreds of \kmsend, and
brightness temperatures of tens of Kelvin
\citep[e.g. ][]{sco97,dow98}.  Similarly, unresolved observations of
starbursts have shown gas and dust temperatures in the range of
30-50 K, in agreement with the models presented here
\citep{yao03,nar05,lee10,mao10,muh11}.

\subsection{Relationship to Other Models}

The seminal work of \citet{mal88} investigated \xco \ in galaxies via
subresolution models of GMCs in a disc-like configuration.  These
authours found that \xco \ would vary from the Galactic value in cases
of high kinetic temperature, high velocity dispersion or low
metallicity. While not simultaneously modeling any of these effects,
this model identified some of the most important driving factors in setting
the observed $X$-factor in clouds.

A number of other studies have also investigated \xco \ in models of
giant molecular clouds.  Early studies implemented 1D radiative
transfer calculations in spherical models of GMCs
\cite[e.g. ][]{kut85,wal07}.  With the increase of computational
power, 3D numerical studies of GMCs in evolution have recently become
feasible. Recently, \citet{glo10} and \citet{glo11} modeled \htwo
\ and CO formation/destruction in magnetohydrodynamic models of GMCs.
These models were elaborated upon by \citet{she11a,she11b} who
utilised radiative transfer calculations in combination with these MHD
models to produce bona fide observables from the model clouds.  These
authours found that model GMC with mean densities, column densities,
temperatures, and velocity dispersions comparable to the Milky Way's
clouds ($n \sim 10^2-10^3 \cmthree$, $N_{\rm H_2} \sim
10^{21}-10^{22} \cmtwo$, $T\sim 10-20 \ {\rm K}$, $\sigma\sim 1-6$ \kms)
had average \xco \ factors of order $2-4\times 10^{20}$\xcounits, and
were insensitive to detailed temperature and velocity distributions.
When manually increasing the velocity dispersion and/or temperature of
the GMC, the resulting \xco \ values fell by a factor of $\sim 5$,
comparable to both observed starbursts and the model mergers in this
paper.  While the simulations in Shetty et al. do not model the
physical processes which may simultaneously cause $N_{\rm H2}$,
$\sigma$, and $T$ to vary, these models do confirm that when
increasing $\sigma$ or $T$ and considering the radiative transfer
through clouds, one will observe a depressed \xco,
as is inferred in ULIRGs.

In this sense, the models of Shetty et al. are complementary to those
presented here.  Shetty et al.'s models resolve much of the physics
and chemistry within GMCs, though they have no information regarding
the external environment from the host galaxy and how it may affect
the cloud.  Our simulations describe the ambient environment
surrounding the model GMCs, though at best they resolve the surfaces of the
clouds (and require some amount of subresolution techniques).  That
both sets of models are converging upon the same result from different
directions is encouraging.  The next step forward will be to fully
couple galaxy evolution simulations with high-resolution models of
GMCs with a grid of model GMCs.  These efforts are underway and will
be presented in due course (R. Feldmann et al. in prep; Narayanan \&
Shetty in prep.).

\section{Summary and Conclusions}
\label{section:summary}

Utilising a combination of hydrodynamic simulations of disc galaxy
evolution and galaxy mergers, dust and molecular line radiative
transfer calculations, we investigated the dependence of the CO-\htwo
\ conversion factor on galactic environment.  Our main results follow:

\begin{enumerate}
\item Provided that GMCs are gravitationally bound, disc galaxies in
  the local Universe have relatively little influence on the physical
  properties of GMCs within them (outside the central $\sim$ kpc).  The
  velocity dispersions are typically dominated by internal processes
  to the GMC, and the temperatures are roughly constant at $\sim10$ K,
  set by a balance of molecular/atomic line cooling and cosmic ray and
  grain photoelectric effect heating.  In this situation, when the
  surface densities of GMCs are comparable to those in the Galaxy,
  \xco \ will be similar to the Galactic value of \xco $ \approx 2-4\times
  10^{20}$ \xcounits.

\item In galaxy mergers, the GMC physical properties are strongly
  affected by the galaxy environment.  The rise in surface density in
  GMCs during the merger is offset by an increase in the velocity
  dispersion coupled to a rise in the kinetic temperature of the gas
  caused by efficient dust-gas thermal exchange at high densities. The
  combination of increased velocity dispersion and kinetic temperature
  increases the CO intensity, and lowers the observed \xco \ from the
  Galactic value by a typical factor of $\sim2-10$.

\item There is a slight trend with galactocentric radius such that
  GMCs toward the centres of disc galaxies will have a lower \xco \ than
  the disc-averaged value, owing to both increased velocity
  dispersions in the clouds, as well as higher kinetic temperatures.
\end{enumerate}

\section*{Acknowledgements} 
This work benefited from discussions had and coding done at the Aspen
Center for Physics.  DN would like thank Patrik Jonsson for numerous
helpful conversations regarding adaptive mesh techniques in radiative
transfer modeling, and Rahul Shetty, Andrew Baker, Emanuele Daddi,
Robert Feldmann, Adam Leroy, Padelis Papadopoulos, and Erik Rosolowsky
for sharing their knowledge on \xco \ in galaxies.  As always, T.J.
Cox's wisdom was invaluable in understanding the physics associated
with galaxy mergers. The authors thank the referee, Simon Glover, for
a constructive report. Finally, DN would like to thank Rob Kennicutt
for providing the original motivation to pursue this study following
an insightful question at the 2007 Gas Accretion and Star Formation
Workshop in Garching.  DN acknowledges support from the NSF via grant
AST-1009452.  MK acknowledges support from: an Alfred P. Sloan
Fellowship; the NSF through grants AST-0807739 and CAREER-0955300; and
NASA through Astrophysics Theory and Fundamental Physics grant
NNX09AK31G and a {\it Spitzer Space Telescope} Theoretical Research
Program grant. ECO acknowledges support from the NSF via grant
AST-0908185.  The simulations in this paper were run on the Odyssey
cluster, supported by the Harvard FAS Research Computing Group.

\newpage

\begin{appendix}
\section{Effects of Parameter Choices and Assumptions}
\label{section:appendix}

In \S~\ref{section:methods}, we outlined a number of parameter choices
which could potentially influence the results in this paper.
Here, we discuss the results in the context of these assumptions.

\subsection{Self-Consistency of The Temperature Calculations}

First, there is a discrepancy between the way the dust temperature is
calculated in \sunrise \ and in our temperature equilibrium model.  In
the former, the dust temperature is assumed to be in equilibrium with
the radiation field, but we do not take into account any thermal
exchange between the gas and dust.  In the temperature equilibrium
model, the dust grains are assumed to be able to exchange energy with
the gas, but we hold the ambient radiation field fixed, rather than
allowing it to change as the dust temperature does.  Given the
importance of the dust temperature in raising the gas temperature in
this model, it is worth investigating any potential differences
between the two dust temperatures.

\begin{figure}
\hspace{-1cm}
\includegraphics[angle=90,scale=0.4]{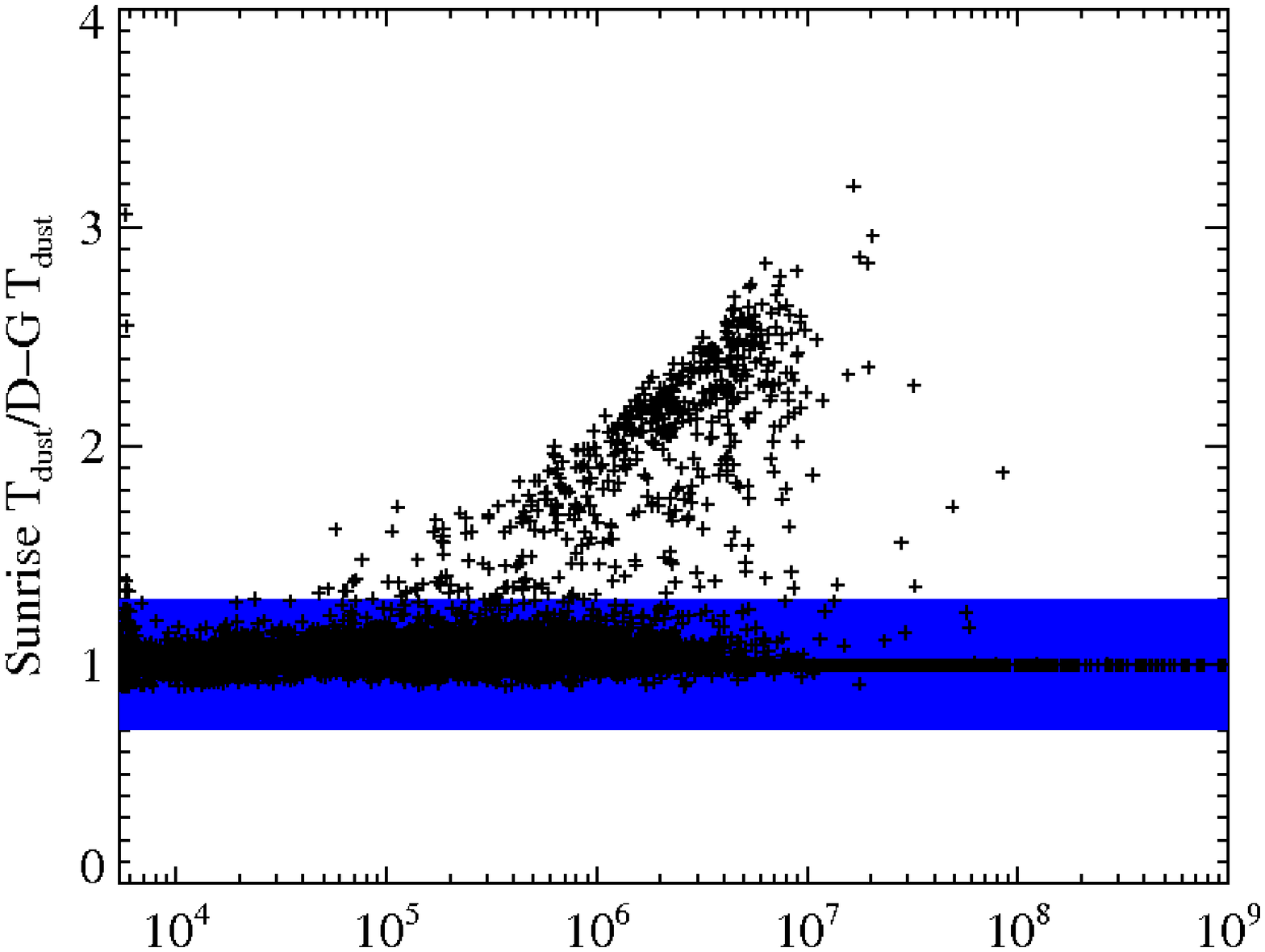}
\vspace{-1cm}
\caption{ Comparison between the \sunrise \ dust
  temperatures and those calculated from the temperature equilibrium
  calculation as a function of density for model GMCs in our fiducial
  merger.  See text for details. \label{figure:dusttemp_contour}}
\end{figure}

In Figure~\ref{figure:dusttemp_contour}, we plot the ratio of $T_{\rm
  dust}$ from \sunrise \ compared to $T_{\rm dust}$ from the
temperature equilibrium model as a function of GMC
density\footnote{This is not the actual mean density of the GMC, but
  the density accounting for an enhancement by the turbulent
  compression of gas.  This is the density that is used in the
  temperature equilibrium calculation.} for our fiducial merger.
There is generally good agreement between the two, though some number
of points at higher densities deviate strongly from unity.  The gas
which has poor agreement between the two dust temperature calculations
almost exclusively has all of its carbon in atomic form, rather than
molecular.  Beyond this, this gas tends to be towards the outskirts of
the galaxy, in rather large cells in the adaptive mesh with relatively
low masses ($\sim 10 \ \msun$).  Because we enforce a rule that clouds
must have a minimum surface density of 100 \msun \ pc$^{-2}$, these
regions have extremely high densities, even if relatively low mass.
In atomic gas of this density, the gas couples with the dust and can
cause the dust temperature to change from that of the background
radiation field.  These outlying points have little effect on the
final results, however, as they contain rather little mass.  We denote
the 95th mass percentile by the blue shaded region in
Figure~\ref{figure:dusttemp_contour}.  That is, the sum of the mass in
the points outside of the blue region accounts for $<5\%$ of the total
molecular mass in the galaxy.  As is shown, the differences in the two
dust temperatures are small in this shaded region.

\subsection{\sunrise \ Input Parameters}

\begin{figure}
\includegraphics[angle=90,scale=0.4]{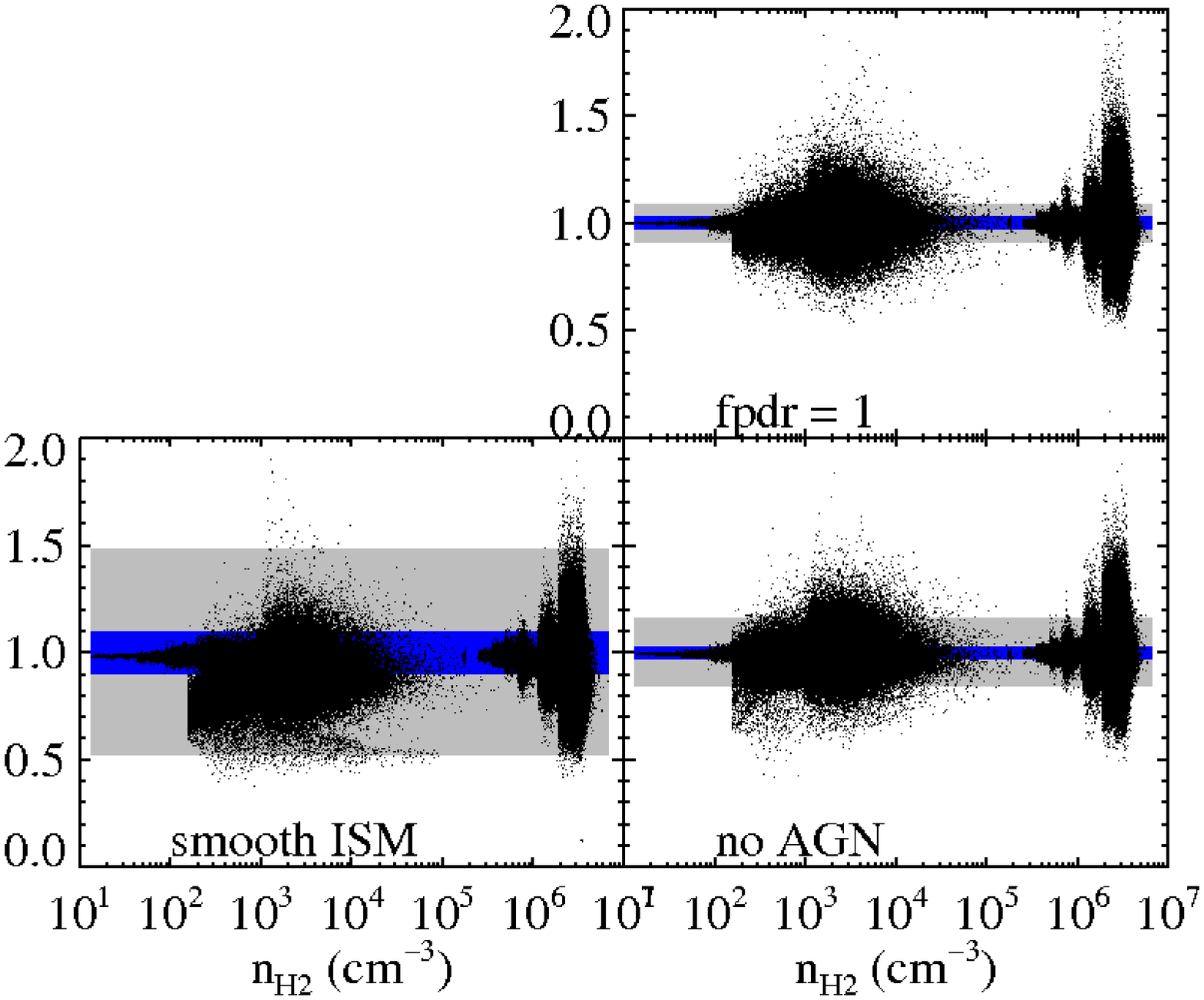}
\caption{Ratio of $T_{\rm dust}$ under various subresolution ISM
  specifications to that derived in our fiducial model versus \htwo
  \ density in model GMCs.  See text for
  details.\label{figure:dusttemp_distribution.lowres}}
\end{figure}

Similarly, a number of our assumptions in the \sunrise \ modeling can
have an effect on the derived dust temperature.  We investigate those
here.  The important figure for these tests is
Figure~\ref{figure:dusttemp_distribution.lowres}.  Referring to
\S~\ref{section:sunrise}, the time-averaged
covering fraction of birthclouds around stellar clusters is a free
parameter.  While we chose a modest covering fraction, it is possible
that a larger fraction may be reasonable.  For example, if we choose a
fraction $f_{\rm PDR} = 1$, the O and B stars would be blanketed by
the ISM for their entire lives.  This situation may exist in ULIRGs.
To test for the effect in varying $f_{\rm PDR}$, we plot the ratio of
the dust temperatures derived from \sunrise \ for a model with $f_{\rm
  PDR} = 1$ over the dust temperature from our fiducial model with
$f_{\rm PDR} = 0.3$ against GMC density.  The blue shaded region shows
the 95th mass percentile, and grey denotes the 99th mass percentile.
The model with a larger clearing timescale for the birthclouds has
cooler dust temperatures as less UV flux interacts with the ISM.
However, the differences in dust temperature are generally within a
factor of 50\%, and much less than that ($<10\%$) when considering the
bulk of the mass of the galaxy.  We conclude that the PDR covering
fraction is not an important driver in our model results.

We can explore the effect of discarding the PDR birthcloud model, and
assuming the cold ISM has a uniform volume filling fraction.  In this
case, the UV photons escape the star particles easily, though optical
depths for the photons in the ISM are large. For some number of the
clouds outside of the nuclear region, the dust temperatures are thus
colder than in our fiducial model (the low ratio points in
Figure~\ref{figure:dusttemp_distribution.lowres}).  However, the bulk
of the gas mass is in a confined nuclear region in the galaxy which
sees the intense UV radiation field.  Because of this, this dust is
heated well, and has comparable dust temperatures to our fiducial
model.  The dispersion in dust temperatures is again within 50\%, and
the 95th\% percentile of mass shows relatively small discrepancies.

It is possible that the AGN in the model merger contributes strongly
to the dust temperature.  To test this, we investigate a model where
we have turned off the contribution of the AGN in determining the dust
temperature in the merger.  While it is difficult to see by eye, the
dust temperatures in the model with the AGN are hotter than the model
with no AGN.  By and large, however, the AGN is not powerful enough to
have a significant effect on the overall temperature structure of the
cold ISM as noted by the blue and grey shaded regions.

\subsection{\gadget \ Input Parameters}

We now turn to possible parameters in the hydrodynamics simulations
which may affect our results. As described in \S~\ref{section:sph}, we
make a number of parameter choices which may affect the star formation
history of the model galaxies.  Because the gas densities and dust and
gas temperature can depend on these assumptions, it is worth exploring
the robustness of our model results in the context of these choices.

There are two principle parameter choices which govern the physical
state of the ISM in our hydrodynamic modes: the star formation
``law'', and the equation of state.  As discussed in
\S~\ref{section:sph}, we adopt a star formation law such that the star
formation time scale is assumed to be proportional to the local
dynamical timescale, and whose rate matches the normalisation of the
locally observed \citet{ken98a} relation.  

In the absence of a complete theory of star formation, a number of
possible choices exist regarding the implementation of a star
formation recipe on subresolution scales.  One can imagine a similar
Kennicutt-Schmidt solution, though with an index of unity as appears
to be observed on resolved scales in nearby galaxies \citep{big08}, or
steeper index as tentative observational \citep{bou07} and theoretical
\citep{fel11} evidence motivates at high-redshift.  Beyond this, more
sophisticated physical models may provide reasonable prescriptions for
star formation in galaxy evolution models
\citep[e.g. ][]{tan00,kru09b,ost10}.  While it is outside the scope of
this study to perform a detailed study of various star formation
recipes in \gadget, we perform some simple tests to investigate the
role of our adopted star formation law in driving the simulated \xco
\ factors.

%\begin{figure}
%\hspace{-1cm}
%\includegraphics[angle=90,scale=0.4]{xco_distributions_appendix.ps}
%\caption{Luminosity-weighted mean \xco for a variety of assumptions
%  in our \gadget \ SPH models.  We refrain from showing the full
%  distributions as the plot becomes unreasonably cluttered.  See main
%  text for details regarding these models. \label{figure:appendix2}}
%\end{figure}

 In the Table, we show the luminosity-weighted mean
 \xco \ values for our fiducial merger during the peak of their
 starburst for a variety of input parameters in our \gadget
 \ simulations.  The Kennicutt-Schmidt index=2 case shows a comparable
 \xco \ value as the fiducial merger.  While the burst is moderately
 diminished (owing to rapid consumption of the gas during early phases
 of the merger's evolution, though still quite large at $\sim 250
 \ \msunyr$ as opposed to $\sim 340 \ \msunyr$ in the fiducial
 merger), the large stellar mass upon coalescence maintains a large
 velocity dispersion in the gas.  This drives a low mean \xco.

The model with the largest impact on our results is the model with
Kennicutt-Schmidt index=1; This model merger has a mean \xco \ comparable to
the model disc.  Models with a KS index of 1 do not go through a burst
upon merging (T. Cox, private communication).  With a KS index of 1,
to first order, the total star formation rate is proportional to the
total gas mass.  Because we don't include any gas replenishment from
the intergalactic medium, the gas mass only decreases with time, as
does the SFR in this simulation.  The low SFR upon merging leads to
low gas/dust temperatures, and increased \xco.  We note that this
situation is unlikely to describe real mergers as, observationally,
galaxy mergers exhibit the highest star formation rates in the local
Universe. Both observational \citep{big08} and theoretical
\citep{kru09b,ost11} evidence suggest that a KS index $> 1$ may
describe high-surface density systems.

We utilise an equation of state for the ISM which incorporates a
subresolution prescription for capturing the effect of supernovae
heating of the ISM \citep{spr03a}.  The nominal \citet{spr03a} and
\citet{spr05a} EOS is given by:

\begin{equation}
P_{\rm eff} = (\gamma-1)(\rho_{\rm h}u_{\rm h} + \rho_{\rm c}u_{\rm c})
\end{equation}
where $\gamma = 5/3$ as the adiabatic index of the gas, $\rho_{\rm
  h,c}$ is the density of the hot and cold phase, and $u_{\rm h,c}$ is
the specific thermal energy of the two phases.  For a given IMF,
\citet{spr03a} show that the EOS is completely defined by the star
formation time scale, the normalisation of the cloud evaporation rate,
and a supernovae ``temperature'' which defines the heating rate from
supernovae of a given IMF.  The full ``effective'' EOS has the
property in which pressure rises with density faster than an
isothermal gas, as can be seen in Figure 4 of \citet{spr05a}.

Our fiducial model utilises a softer EOS than the full model.  In
particular, we interpolate between the full ``stiff'' model (where we
assign a parameter $q_{\rm EOS}=1$) and isothermal model ($q_{\rm
  EOS}=0$) and employ $q_{\rm EOS}=0.25$. In the full ``stiff'' EOS as
in simulations of discs scaled for the local Universe, the ISM can
become so pressurised as to appear smooth with relatively few clumps
\citep{spr05a}. To test our assumption of a softer EOS, we have run a
test simulation with \qeos=1.

In the Table, we show the mean \xco \ for our fiducial merger, though
with \qeos=1 (denoted ``Stiff EOS''). We see a larger value than the fiducial model. Isolating
the root cause is nontrivial. By effectively increasing the effect of
supernovae feedback, we increase dust heating (by reducing the
clumpiness of the gas), though we also reduce the magnitude of the
burst (due to a retardation of gas fragmentation).  These effects
serve to somewhat offset one another with respect to the gas
temperature.

We test whether our assumption of $G_0 = 1$ outside of clouds plays a
strong role in our model results.  Following \citet{ost10}, we have
run a model in which we scale the interstellar radiation field by the
value of the local SFR compared to that in the solar neighbourhood,
and show the mean \xco \ value in the Table (denoted
by ``OML G-Scaling''). Because the clouds are strongly shielded in
this model, scaling $G$ makes little difference.

Finally, we consider the spatial resolution of our model.  Our current
resolution has cell sizes ($\sim 70$ pc) which are of order the SPH
smoothing length.  Further increasing the spatial resolution does not
provide new physical information.  More seriously, increasing the
spatial resolution of the SPH simulations would run into scenarios of
unphysical descriptions of the ISM with the \citet{spr03a} multiphase
model.  We can see, however, what direction the results may go if it
were possible to increase our resolution.  To investigate this, we
have run our fiducial merger snapshot with one less level of
refinement in the adaptive mesh, giving a minimum cell size of $\sim
140 $pc, and show the mean \xco \ in the Table (denoted by
``Half-Res'').  The mean \xco \ in the low resolution model is 60\%
\ the value of the fiducial model.  It is conceivable, then, that
further increasing the spatial resolution would increase \xco \ in the
merger model. On the other hand, reducing the cell size would result
in more resolved GMCs. Because it is the resolved GMCs that drive the
mean \xco \ for the merger model, it is also possible that increased
spatial resolution would result in little change to the results
presented here.

\begin{table}
\label{table:appendix}
\centering
\begin{minipage}{100mm}
\caption{}
\begin{tabular}{@{}cc@{}}
\hline Model & Mean \xco \\
&(\xcounits)\\
\hline
Fiducial & $6.31 \times 10^{19}$ \\
Stiff EOS & $1.08 \times 10^{20}$\\
KS = 2 & $5.43 \times 10^{19}$\\
KS = 1 & $2.73 \times 10^{20}$ \\
OML G-scaling & $7.10 \times 10^{19}$\\
Half-Res & $3.02 \times 10^{19}$\\
\hline
\end{tabular}
\end{minipage}
\end{table}
\end{appendix}

\end{document}